%% file: bruzual.tex
\documentclass[11pt,twoside]{article}
\usepackage{cozumel2005}
\usepackage{epsf}
\usepackage{psfig}
\usepackage{lscape}
\begin{document}
\title{Studying stellar populations at high spectral resolution}
\author{Gustavo Bruzual A.}
\affil{C.I.D.A., Apartado Postal 264, M\'erida 5101-A, Venezuela}

\begin{abstract}
I describe very briefly the new libraries of empirical spectra of stars
covering wide ranges of values of the atmospheric parameters T$_{eff}$,
log g, [Fe/H], as well as spectral type, that have become available in the
recent past, among them the HNGSL, MILES, UVES-POP, ELODIE, and the IndoUS
libraries. I show the results of using the IndoUS and the HNGSL libraries,
as well as an atlas of theoretical model atmospheres, to build population
synthesis models. These libraries are complementary in spectral resolution
and wavelength coverage, and will prove extremely useful to describe spectral
features expected in galaxy spectra from the NUV to the NIR.
The fits to observed galaxy spectra using simple and composite stellar population
models are discussed.
\end{abstract}

\section{Introduction}

\citet{BC03}, hereafter BC03, have examined in detail the
advantages of using intermediate resolution stellar spectra in population
synthesis and galaxy evolution models. In these models BC03 use the STELIB
library compiled by \citet{STELIB03}.
This library contains observed spectra of 249 stars in a wide range of
metallicities in the wavelength range from 3200 \AA\ to 9500 \AA\ at a
resolution of 3 \AA\ FWHM (corresponding to a median resolving power of
$\lambda / \Delta\lambda \approx 2000$), with a sampling interval of 1 \AA\ 
and a signal-to-noise ratio of typically 50 per pixel.
The BC03 models reproduce in detail typical galaxy spectra extracted from the
SDSS Early Data Release, e.g., \citet{EDR02}.
From these spectral fits one can constrain physical parameters such as the
star formation history, metallicity, and dust content of galaxies, e.g.,
\citet{HPJD04}; \citet{CF05}; \citet{MAT05}.
The medium resolution BC03 models also enable accurate studies of
absorption-line
strengths in galaxies containing stars over the full range of ages and can
reproduce simultaneously the observed strengths of those Lick indices that
do not depend strongly on elemental abundance ratios, provided that the
observed velocity dispersion of the galaxies is accounted for properly,
and offer the possibility to explore new indices over the full optical range
of the STELIB atlas, i.e. 3200 \AA\ to 9500 \AA.
To extend the spectral coverage in the models beyond these limits, BC03
recur to other libraries.
For solar metallicity models, the \citet{PIC98} library can be used to extend
the STELIB spectral coverage down to 1150 \AA\ in the UV end and up to 2.5
$\mu$m at the red end, with a sampling interval of 5 \AA\ pixel$^{-1}$ and a
median resolving power of ($\lambda / \Delta\lambda \approx 500$).
The UV spectra in the \citet{PIC98} atlas are based on $IUE$ spectra of bright
stars.
At all metallicities the BC03 models are extended
down to 91 \AA\ in the UV side and up to 160
$\mu$m at the other end using the theoretical model atmospheres included
in the BaSeL series of libraries compiled by \citet{LEJ97, LEJ98}, and
\citet{WES02}, but at a resolving power considerably lower than for STELIB
($\lambda / \Delta\lambda \approx 200-500$).

Given the success in reproducing observed galaxy spectra with the BC03 
synthesis models, it is important to explore models built with
libraries of higher spectral resolution and which improve upon STELIB
on the coverage of the HRD by including at all metallicities
a broader and more complete distribution of spectral types.
This goal is now possible thanks to several compilations of stellar spectra
that have become available in the last few years.
In this paper I describe briefly the properties of these libraries and
show applications of galaxy models built using some of them.
The full implementation of the new libraries in the population synthesis
models is in preparation by Bruzual \& Charlot.

\input fig1

\section{Stellar Libraries}

A relatively large large number (5) of libraries containing medium to
high spectral resolution observed spectra of excellent quality for large
numbers of stars (hundreds to thousand) have become available
during the recent past. One of the main objectives of the observers
who invested large amounts of time and effort assembling these data sets was
to build libraries suitable for population synthesis. In this respect,
the stars in the libraries have been selected to provide broad coverage of
the atmospheric parameters T$_{eff}$, log g, [Fe/H], as well as spectral type,
throughout the HRD.
In parallel to this observational effort, considerable progress has been
made in the computation of theoretical model atmospheres at high spectral
resolution for stars whose physical parameters are of interest for galaxy
modeling.
Below I describe very briefly the characteristics of each of these libraries
which are relevant for population synthesis models.

\subsection{HNGSL}

The Hubble's New Generation Spectral Library \citep{HL03} contains spectra
for a few hundred stars whose fundamental parameters, including chemical
abundance, are well known from careful analysis of the visual spectrum.
The spectra cover fully the wavelength range from 1700 \AA\ to 10,200 \AA.
The advantage of this library over the ones listed below is the excellent
coverage of the near-UV and the range from 9000 \AA\ to 10,200
\AA, which is generally noisy or absent in the other data sets.

\subsection{MILES}

The Medium resolution INT Library of Empirical Spectra \citep{MILES05},
contains carefully calibrated and homogeneous quality spectra for 1003 stars
in the wavelength range 3500 \AA\ to 7500 \AA\ with 2 \AA\ spectral resolution
and dispersion 0.9 \AA\ pixel$^{-1}$. The stars included in this library were
chosen aiming at sampling stellar atmospheric parameters as completely as
possible.

\subsection{UVES POP library}

The UVES Paranal Observatory Project \citep{UVES03}, has produced a library
of high resolution ($R = \lambda / \Delta\lambda \approx 80,000$) and high
signal-to-noise ratio spectra for over 400 stars distributed throughout
the HRD. For most of the spectra, the typical final SNR obtained in the V band
is between 300 and 500. The UVES POP library is the richest available database
of observed optical spectral lines.

\subsection{IndoUS library}

The IndoUS library \citep{VAL04} contains complete spectra over the entire
3460 \AA\ to 9464 \AA\ wavelength region for 885 stars obtained with the
0.9m Coud\'e Feed telescope at KPNO. The spectral resolution is $\approx$
1 \AA\ and the dispersion 0.44 \AA\ pixel$^{-1}$. The library includes
data for an additional 388 stars, but only with partial spectral coverage.

\subsection{ELODIE}

The ELODIE library is a stellar database of 1959 spectra for 1503 stars,
observed with the \'echelle spectrograph ELODIE on the 193 cm telescope at
the Observatoire de Haute Provence. The resolution of this library is 
$R = 42,000$ in the wavelength range from 4000 \AA\ to 6800 \AA\ 
\citep{PAS01A, PAS01B}.
This library has been updated, extended, and used by \citet{LeB04}
in the version 2 of the population synthesis code PEGASE.

\subsection{High-spectral resolution theoretical libraries}

There are several on-going efforts to improve the existing grids of
theoretical
model atmospheres including the computation of high resolution theoretical
spectra for stars whose physical parameters are of interest for population
synthesis. See, for example, \citet{COE05}; \citet{BER04}; \citet{MM04};
\citet{PET04}, as well as the papers by Ch\'avez, Bertone, Buzzoni \&
Rodr{\'\i}guez-Merino; Gonz\'alez-Delgado \& Cervi\~no; and Munari \&
Castelli in this conference.

\section{Using the New Libraries in Population Synthesis Models}

In BC03 the 'standard' reference model represents a simple stellar
population (SSP) computed using the Padova 1994 evolutionary tracks,
the \cite{CHAB03} IMF truncated at 0.1 and 100 M$_\odot$, and either
the STELIB, the Pickles, or the BaSeL 3.1 spectral libraries
(see BC03 for details).
Below I show the behavior of the standard reference model computed with
the IndoUS and the HNGSL libraries, and the \citet{COE05} atlas
of theoretical model atmospheres.

\input fig2
\input fig3
\input fig4
\input fig5

\subsection{IndoUS library}

Figure 1 shows a histogram of the number of stars per 0.05 size bin
in [Fe/H] for both the IndoUS and the STELIB libraries. The gain in the
number of stars available at most metallicities in the IndoUS library
with respect to STELIB shows clearly in this plot.
This should translate in a better sampling of most stellar types at
the relevant positions in the HRD for the metallicities corresponding
to the different evolutionary tracks, also shown in Figure 1.
Models built using the IndoUS library are compared to the STELIB and BaSeL
3.1 models in Figures 2 to 5. As expected, spectral features are more
numerous and more clearly seen in the IndoUS model than in the other two models.
Colors measured from the continuum flux in these three spectra are all very similar.

Figures 6 and 7 show the normalized residual distributions resulting from
simultaneous fits of the strengths of several indices in the spectra
of 2010 galaxies with S/N$_{\rm med} \ge 30$ in the `main galaxy sample'
of the SDSS EDR using BC03 models built with the STELIB and the IndoUS libraries,
respectively.
The highlighted frames indicate the seven indices used
to constrain the fits. The stellar velocity dispersion of the models are
required to be within 15 km s$^{-1}$ of the observed ones. Each panel shows
the distribution of the fitted index strength $I^{\rm fit}$ minus the
observed one $I^{\rm obs}$, divided by the associated error $\sigma_I$.
For reference, a dotted line in each panel indicates a Gaussian distribution
with unit standard deviation. The shaded histograms show the contributions
to the total distributions by galaxies with $\sigma_V >180$ km s$^{-1}$,
corresponding roughly to the median stellar velocity dispersion of the
sample. Figure 6 is identical to Figure 18 of BC03.
It is apparent from a comparison of these two figures that the CN$_1$, CN$_2$,
Ca4227, G4300, TiO$_1$, and TiO$_2$ indices measured in this
galaxy sample are reproduced more closely by the IndoUS models than the
STELIB models. The Mg$_1$, Mg$_2$, Mgb, and NaD indices are equally off
in both sets of models. The negative residuals indicate that the model
values are below the observed ones, due to the lack of stars with 
enhanced $\alpha$ element abundance in these stellar libraries.
The distributions for H$\beta$ and H$\delta_A$ in the IndoUS models
are shifted toward positive and negative residual values, respectively,
with respect to the STELIB models. The four [MgFe] indices shown
in the bottom row of Figures 6 and 7 are all shifted toward negative
residuals in the IndoUS models.

\input fig6
\input fig7
\input fig8
\input fig9
\input fig10

\subsubsection{Flux calibration problems in the IndoUS library.}

In Figure 8 I compare the spectra of 6 of the more than 500 stars in
common between the IndoUS and the MILES libraries.
These plots illustrate a flux calibration problem affecting some
of the IndoUS spectra.
Figure 9 shows spectra of 6 of the stars in common in the IndoUS, the MILES,
and the STELIB libraries.
For all the stars in common in the three libraries, the shape of the MILES
and STELIB spectra always match, whereas if there is a discrepancy it can
always be attributed to the IndoUS spectrum, as in the case of the star
HD122563 in the upper right hand side panel of Figure 9.
The IndoUS spectra of stars with flux calibration problems should not
be used in population synthesis unless they can be re-calibrated correctly.
In Figure 10 I indicate a procedure that uses the BaSeL 3.1 spectrum
for the values of $T_{eff}$, log g, and [Fe/H] of the problem star
to define the correct shape of the IndoUS spectrum.
This procedure assumes that the physical parameters of the star are
known accurately.

\subsection{HNGSL}

I have also computed the BC03 standard model using the HNGSL \citep{HL03}
instead of STELIB to represent the stellar spectra.
Figure 11 shows clearly the advantages of using the HNGSL to study
the near UV below 3300 \AA. Spectral features, lines and discontinuities,
are much better defined in the HNGSL spectra than in the IUE spectra (used in the
Pickles library) and the BaSeL 3.1 atlas (see BC03 and \citet{GB04} for details).
However, colors measured from the continuum flux in these three spectra
are very similar.
The higher spectral resolution of STELIB above 3300 \AA\ compared to
HNGSL is clearly seen in this figure.

The spectra in Figure 12 are useful to study the different behavior of
models computed with different libraries in the region around and above
9000 \AA. There is a clear difference in the continuum level predicted
by the HNGSL and STELIB models (lowest level) compared to the level predicted
by the BaSeL 3.1 and the Pickles library models (highest level).
The HNGSL model is to be preferred, given the higher SNR of this spectrum
in this region. However, spectral features below 8500 \AA\ are more clearly
seen in the higher resolution STELIB model.

\subsection{Theoretical model atmospheres}

Figures 13 and 14 show the BC03 standard SSP model sed at 12 Gyr
computed with the high resolution theoretical model atmospheres from
the IAG collaboration for [Fe/H] = 0, $[\alpha/Fe] = 0$, in the wavelength
range from 0.3 to 18 $\mu$m \citep{COE05}.
In Figure 14 one can clearly see those regions of the spectrum where the
theoretical atmosphere models produce different results from the empirical
library models, e.g., above 9000 \AA.
In Figure 15 I compare the STELIB and IndoUS models with the IAG collaboration
model degraded in spectral resolution to match the empirical spectra.
This figure shows a remarkable degree of agreement between the empirical
and theoretical models when the difference in resolution is duly taken into
account. This figure replaces the one I showed in my oral presentation which
contained an involuntary error in the procedure used to downgrade the
spectral resolution of the theoretical model.

\input fig11
\input fig12
\input fig13
\input fig14
\input fig15

\section{Reproducing Galaxy SEDs}

\subsection{SSP fits}

By means of a standard least-squares technique it is possible to select
the age and stellar velocity dispersion at which a given SSP model
reproduces most closely a given observed galaxy spectrum.
Figures 16 to 18 show remarkably good fits to the continuum spectrum of three
galaxies with different ages and velocity dispersions, and
in different wavelength intervals. 
The BC03 standard SSP model computed with the IndoUS stellar library
was used in these fits. The stellar velocity dispersion $\sigma$ indicated
in each figure and applied to the model spectrum is also derived by the
fitting algorithm.
The residuals (observed - model) shown at the bottom of the plots are
quite flat over the whole spectral range.

\subsection{Non-parametric CSP fits}

An alternative approach for studying the stellar populations present in a galaxy, 
based on the fact that galaxies are thought to be more closely described by
a composite stellar population (CSP) rather than by an SSP, consists in deriving
the galaxy star formation history (SFH) by means of a non-parametric CSP fit to
the galaxy sed. In a non-parametric fit the SFH is not assumed to be described
by an analytic function, e.g., an exponentially decaying star formation rate
with characteristic $e-$folding time $\tau$.
Instead, all the spectra defining the evolution of an SSP of one or more metallicity
values are allowed to be selected individually by a spectral fitting algorithm.
The SFH for the galaxy is rebuilt from the known age and metallicity of each of the
selected spectra. See \citet{MAT05} for a description of GASPEX, a non-parametric
CSP fitting procedure based on the {\it Non-Negative Least Squares} (NNLS)
algorithm of \citet{LH74}.

The SFH recovered by this type of algorithm is highly dependent on the
signal-to-noise ratio and the wavelength range covered by the problem spectrum.
Figures 19 and 20 show the results of applying the GASPEX algorithm to a problem
spectrum of known SFH. The problem spectrum corresponds to an old stellar
population of age 6 Gyr in which a second burst of star formation of $1 \over 9$
the intensity of the first burst occurs 360 Myr ago. In these figures, the two
vertical lines at the corresponding ages represent schematically this SFH.
The histograms with error bars in Figure 19 represent the SFH recovered by the
GASPEX algorithm when noise in varying amounts is added to the problem spectrum.
The value of the signal-to-noise ratio (SNR) used is indicated in each frame.
Values of SNR above 10 are required to recover the input SFH.
Similarly, Figure 20 shows that the broader the wavelength range covered by the
problem spectrum, the more likely it is that we recover the input SFH. For the
example in Figure 20, the problem spectrum is the same one used in Figure 19
for a signal-to-noise ratio of 20. The full coverage of the wavelength range
from the UV to the IR is required to recover the known SFH.

Figure 21 shows the SFH and cumulative mass vs. age function recovered by the
GASPEX algorithm for a selection of galaxies from the Near Field Galaxy Survey
\citep{JAN00}.
The thin lines represent galaxies with significant recent star formation,
as indicated by the emission lines in their spectra.
It is apparent from this plot that these galaxies assemble their stars at
a later time than the early-type galaxies represented by the heavy lines in
Figure 21.

\subsection{SSP vs. CSP fitting}

Figure 22 shows the best SSP model fit to the spectrum of an early type
galaxy resulting from co-adding the spectra of several galaxies from the
SDSS until reaching a signal-to-noise ratio of 250. This spectrum, kindly
provided by Jarle Brinchmann, is shown as a gray line in the wavelength
range from 3600 to 7500 \AA.
The BC03 standard SSP model computed with the STELIB stellar library for
solar metallicity at an age of 9.25 Gyr is shown in black.
The stellar velocity dispersion $\sigma = 280$ km s$^{-1}$ applied to the
model spectrum is also derived by the fitting algorithm.
At the bottom of the figure the line marked SSP represents the residuals
(observed -model) for the SSP model fit shown in the figure.
The line marked CSP represents the residuals obtained when the fit is
performed by the non-parametric GASPEX CSP fitting algorithm described by
\citet{MAT05}. The line marked S-C shows the difference between the SSP and
CSP solutions (shifted down arbitrarily for clarity).

The top frame of Figure 23 shows the cumulative squared residuals ($\propto \chi^2$)
for the SSP and CSP fits shown in Figure 22.
The CSP fit reaches a lower $\chi^2$ value. The improved
fit over the SSP fit is achieved by the GASPEX algorithm by assuming that some
star formation happened in this galaxy at ages younger than the SSP age of 9 Gyr.
The vertical lines show the position of the central bands defining
the CN$_1$-CN$_2$, Mg$_1$-Mg$_2$-Mgb, and NaD Lick indices, and the H$\alpha$ line.
The lower residuals in the CSP fit are possible because the younger population
fills in the deficiency of the SSP model at the bands containing the Mg and Na
features, i.e. the steep increase in $\chi^2$ seen at these wavelengths in the
SSP fit are not present in the CSP fit. On the other hand, the CN features
are reproduced best by the SSP model. The H$\alpha$ emission line is not included
in neither of the models. The same is true for the apparent features in the observed
spectrum at $\lambda > 6700$ \AA. Beyond H$\alpha$ the two $\chi^2$ functions
run almost parallel.

The middle frame of figure 23 shows the percentage contribution of the old and a very
young population (100 Myr) to the total spectrum of this galaxy in the CSP
solution as a function of wavelength.
The contribution of populations of other ages included in the
GASPEX solution is, added all together, below the 5\% level at
$\lambda < 4500$ \AA, and is even less at longer wavelengths.
The spectra corresponding to the old, young, and the rest of the stellar populations
in the GASPEX solution are shown in the bottom frame of this figure.
The latter contribute essentially zero flux to the total sed. The heavy
line in this frame represents the CSP solution, i.e., the addition of the
individual spectra.

The fact that the best GASPEX solution has a lower $\chi^2$ and requires
recent star formation does not necessarily imply that it represents a more
realistic solution than the SSP fit from the astrophysical point of view.
However, the GASPEX non-parametric CSP solution for this galaxy es reminiscent
of the E+A phenomenon, quite common in E galaxies.
The fact that the CSP solution seems able to detect the E+A phenomenon
makes this technique appealing.

Figure 24 shows a detailed comparison of the fit to one of J. Brinchmann co-added
spectra in specific wavelength regions. The residuals in the second, fourth and sixth
panels are expressed in units of the standard deviation $\sigma$ of the co-added fluxes.
The signal-to-noise ratio in these spectra is so high, and $\sigma$ so low, that
the very small differences between the problem and solution spectra translate into
very large residuals when measured in units of $\sigma$. Thus, very careful analysis
of the fits should be performed when deriving galaxy SFHs.

\section{Conclusions}

New libraries of empirical spectra of stars covering wide ranges
of values of the atmospheric parameters T$_{eff}$, log g, [Fe/H],
as well as spectral type, that have become available recently,
are complementary in spectral resolution and wavelength coverage.

Models built using the IndoUS and the HNGSL empirical stellar
libraries, and the IAG collaboration theoretical model atmospheres,
show that this sort of library will prove extremely useful in describing
spectral features expected in galaxy spectra of various ages
and metallicities from the NUV to the NIR.

Models build with the IndoUS library increase the number of Lick
indices that are correctly reproduced by the models.

The use of population synthesis models to derive galaxy star formation
histories by fitting galaxy spectra have become very popular in recent years.
The reader should keep in mind that different solutions are provided by
the simple minimum $\chi^2$ SSP fit and the more complex non-parametric
CSP fit. Even if the CSP fits result in a lower value of $\chi^2$, the
solutions are not necessarily more sound from astrophysical grounds.
The fact that the CSP solution seems able to detect the E+A phenomenon
makes this technique appealing.

Complete sets of models that use the libraries mentioned in \S2
are been built and will be discussed in a coming paper by Bruzual \& Charlot.

\acknowledgements

I thank Jarle Brinchmann for providing me with his sample of co-added
SDSS spectra.

\input fig16
\input fig17
\input fig18
\input fig19
\input fig20

\input bibliography
\input fig21
\input fig22
\input fig23
\input fig24

\end{document}

%% file: fig1.tex
\begin{figure}
\centerline{\vbox{\psfig{figure=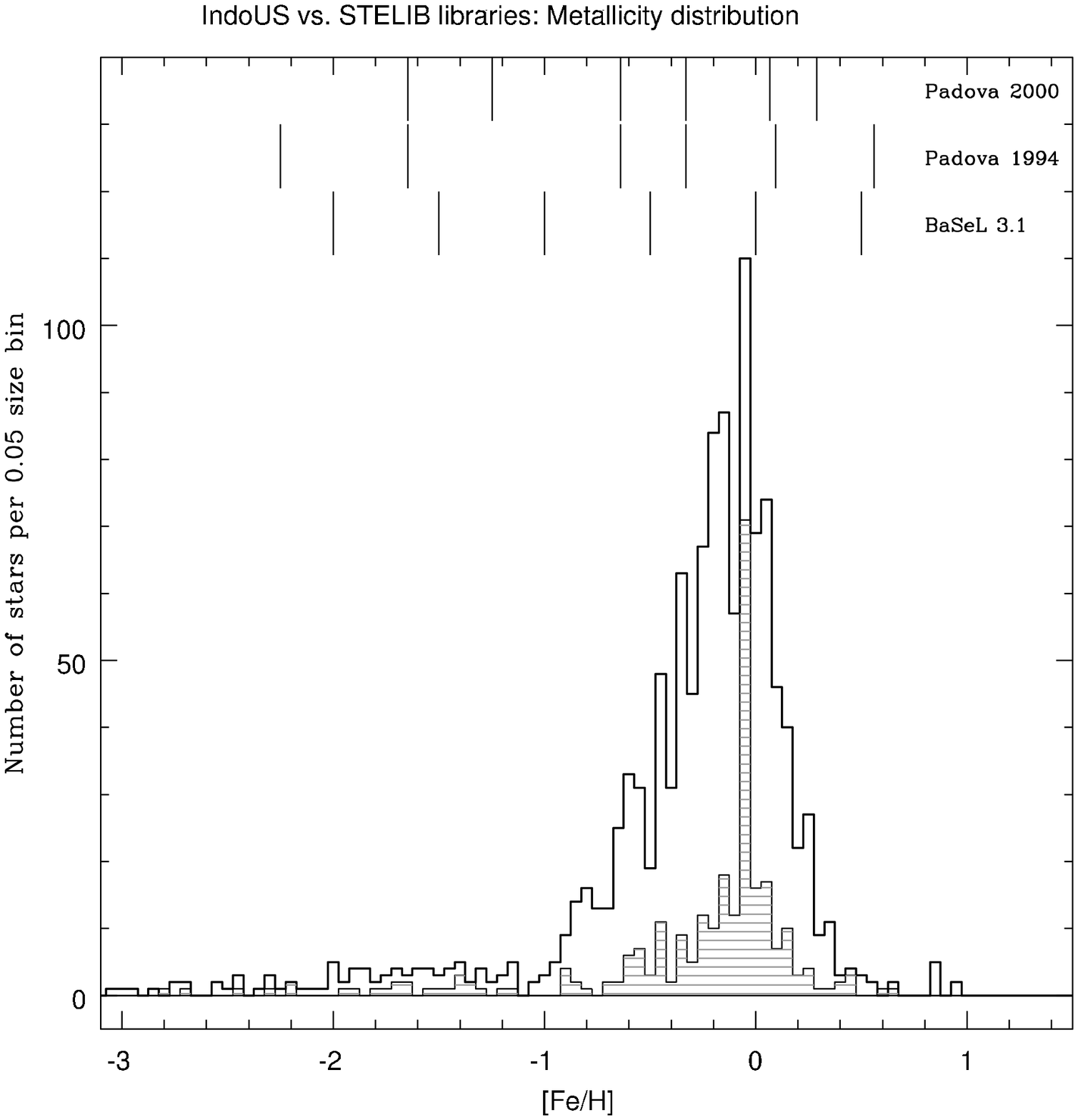,width=14cm,height=20cm,angle=0}}}
\vskip -4.00cm
\caption{
  Histogram showing the number of stars per 0.05 size bin in [Fe/H] for
  the IndoUS (dark line) and STELIB (hatched histogram) libraries. The values
  of [Fe/H] corresponding to the Padova 1994 and Padova 2000 sets of
  evolutionary tracks and the BaSeL 3.1 spectral atlas are shown as
  vertical lines at the top of the figure.
}
\end{figure}

%% file: fig2.tex
\begin{figure}
\centerline{\vbox{\psfig{figure=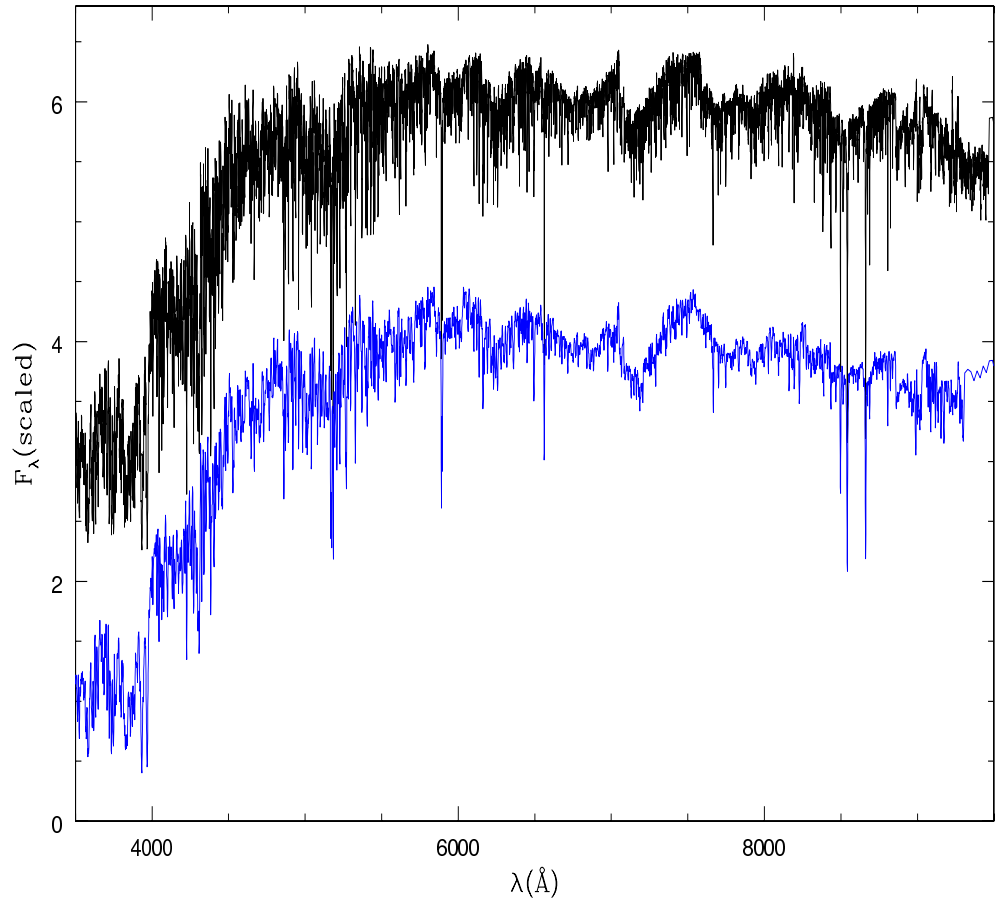,width=14cm,height=13cm,angle=270}}}
\caption{
  BC03 standard SSP model spectra for solar metallicity at 12 Gyr in
  the wavelength range from 3500 \AA\ to 9500 \AA\ computed with the
  IndoUS (top line) and the STELIB (bottom line) libraries. The spectra
  have been scaled and shifted in the vertical direction for clarity.
}
\end{figure}

%% file: fig3.tex
\begin{figure}
 \setbox20=\hbox
 {\psfig{file=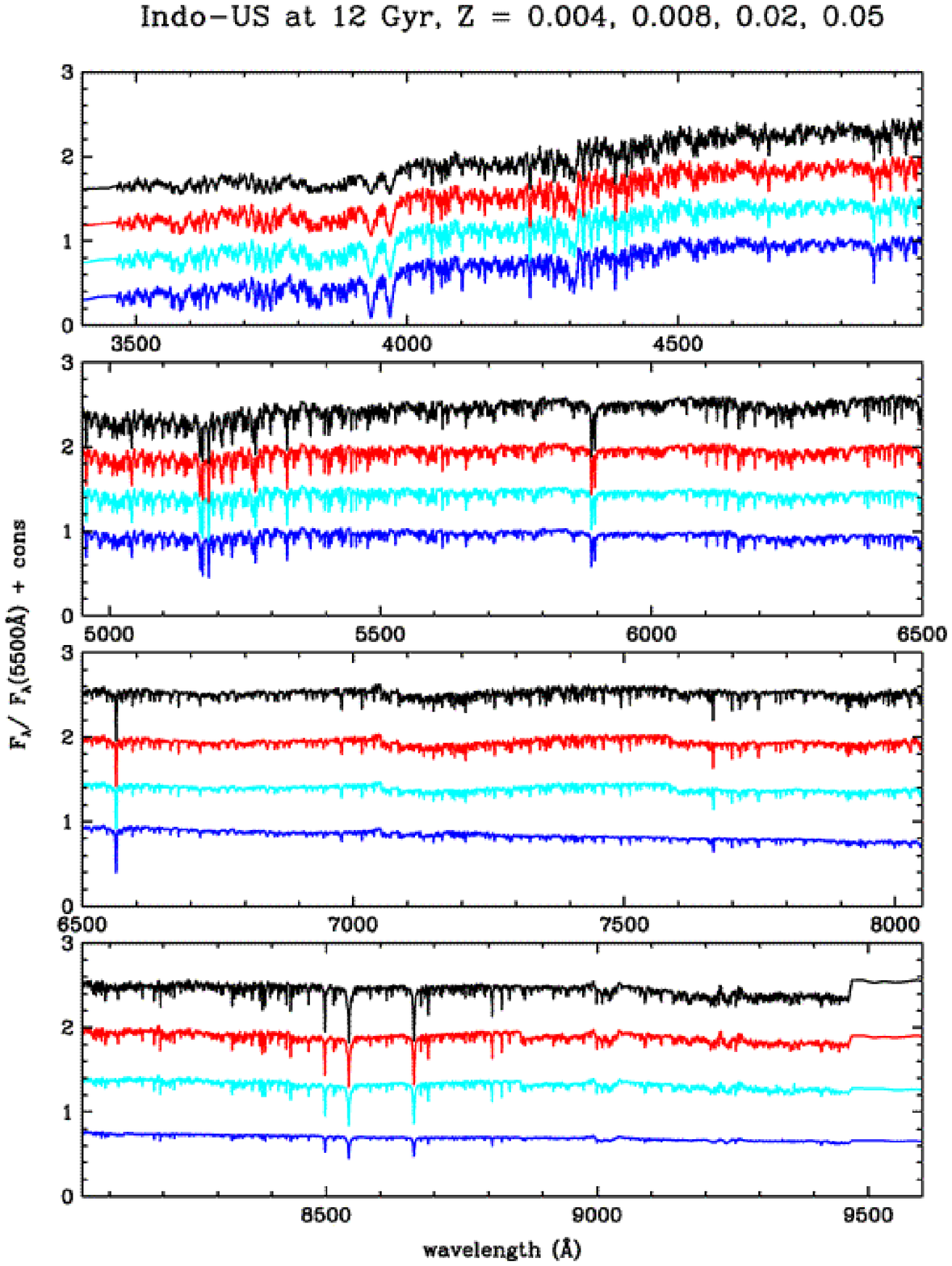,width=14cm,height=20cm,angle=0}}
 \centerline{\box20}
 \caption{
  BC03 standard SSP model spectra for Z = 0.004, 0.008, 0.02, and 0.05 (top
  to bottom) at 12 Gyr in the wavelength range from 3500 to 9500 \AA\ 
  built using the IndoUS stellar library. The spectra have been
  normalized and shifted arbitrarily in the vertical direction for clarity.
}
\end{figure}

%% file: fig4.tex
\begin{figure}
 \setbox20=\hbox
 {\psfig{file=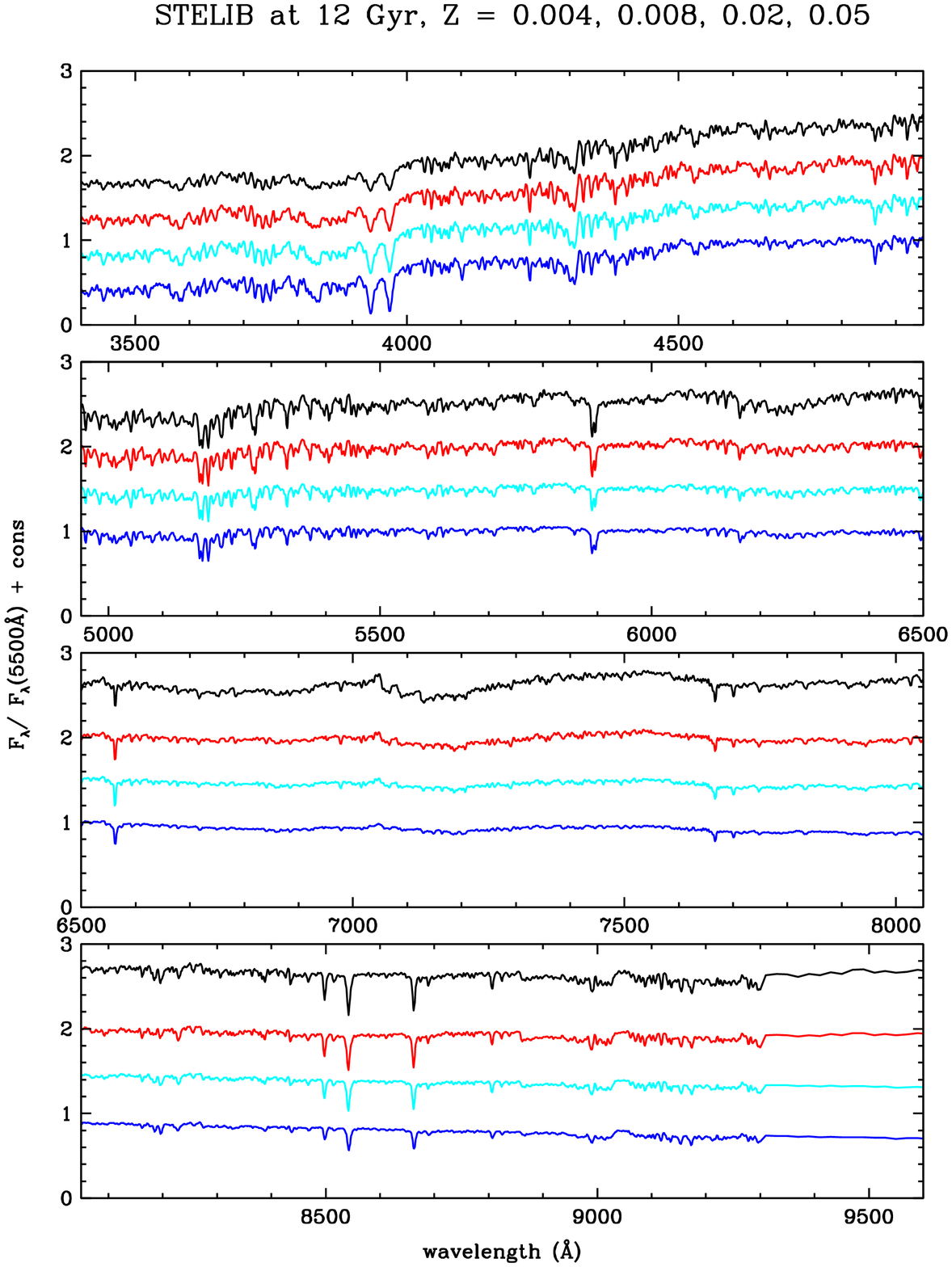,width=14cm,height=20cm,angle=0}}
 \centerline{\box20}
 \caption{
 BC03 standard SSP model spectra for Z = 0.004, 0.008, 0.02, and 0.05 (top
  to bottom) at 12 Gyr in the wavelength range from 3500 to 9500 \AA\
  built using the STELIB stellar library. The spectra have been
  normalized and shifted arbitrarily in the vertical direction for clarity.
}
\end{figure}

%% file: fig5.tex
\begin{figure}
 \setbox20=\hbox
 {\psfig{file=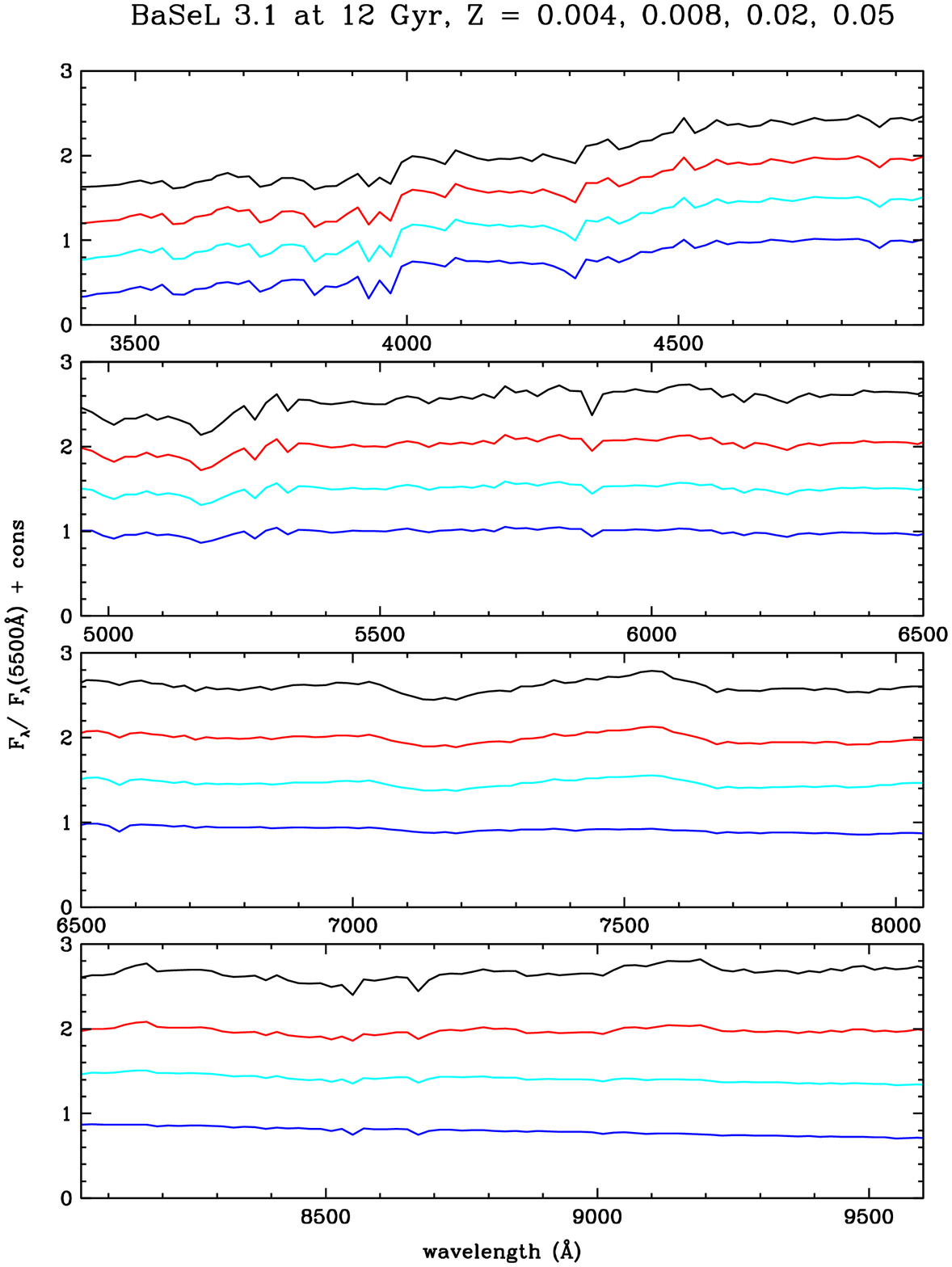,width=14cm,height=20cm,angle=0}}
 \centerline{\box20}
 \caption{
 BC03 standard SSP model spectra for Z = 0.004, 0.008, 0.02, and 0.05 (top
  to bottom) at 12 Gyr in the wavelength range from 3500 to 9500 \AA\
  built using the BaSeL 3.1 spectral atlas. The spectra have been
  normalized and shifted arbitrarily in the vertical direction for clarity.
}
\end{figure}

%% file: fig6.tex
\begin{figure}
 \setbox20=\hbox
 {\psfig{file=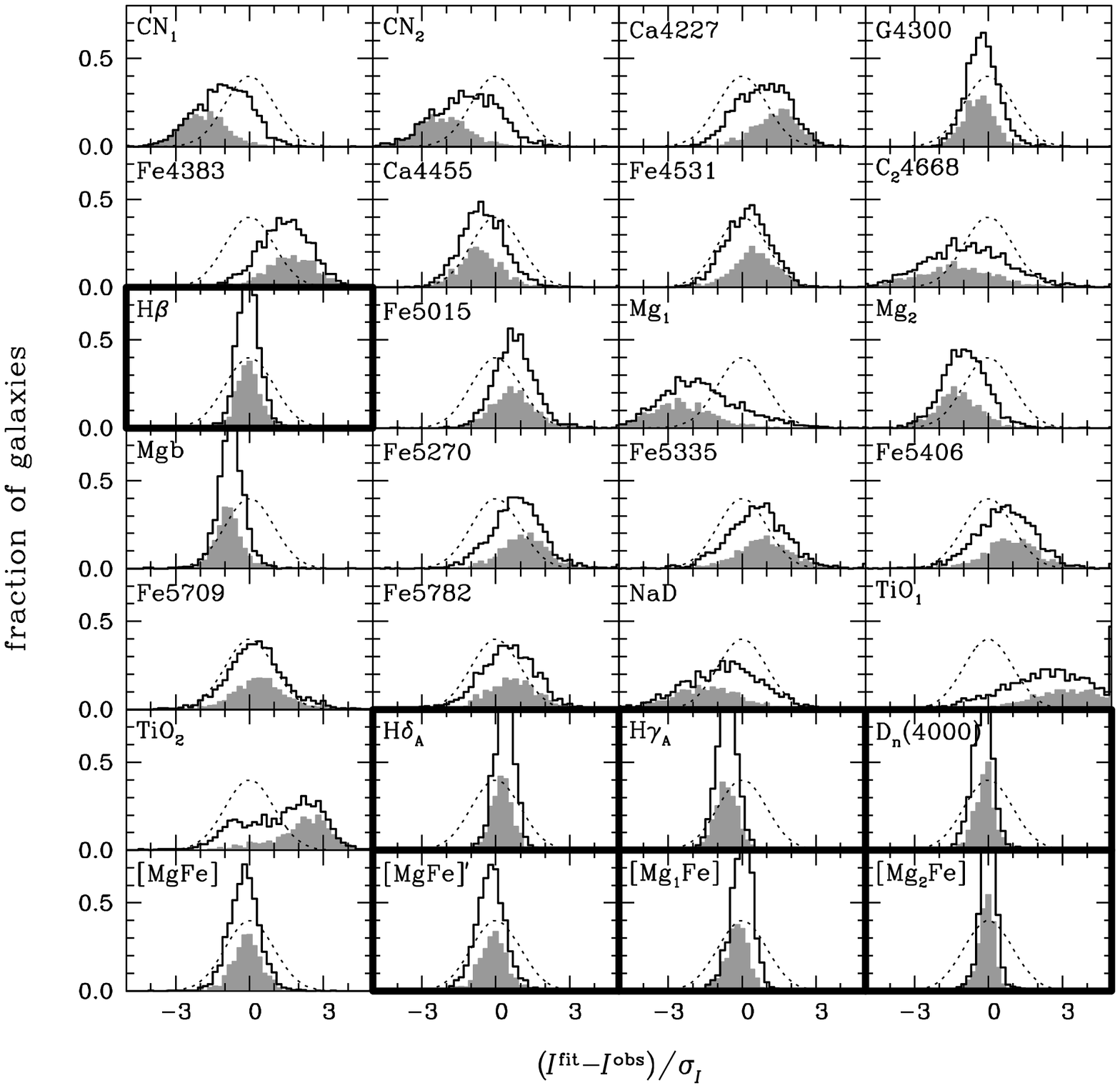,width=14cm,height=20cm,angle=0}}
 \centerline{\box20}
 \caption{
  Simultaneous fit of the strengths of several indices in the spectra
  of 2010 galaxies with S/N$_{\rm med} \ge 30$ in the `main galaxy sample'
  of the SDSS EDR using BC03 models built with the STELIB library.
  The highlighted frames indicate the seven indices used
  to constrain the fits. The stellar velocity dispersion of the models are
  required to be within 15 km s$^{-1}$ of the observed ones. Each panel shows
  the distribution of the fitted index strength $I^{\rm fit}$ minus the
  observed one $I^{\rm obs}$, divided by the associated error $\sigma_I$.
  For reference, a dotted line in each panel indicates a Gaussian distribution
  with unit standard deviation. The shaded histograms show the contributions
  to the total distributions by galaxies with $\sigma_V >180$ km s$^{-1}$,
  corresponding roughly to the median stellar velocity dispersion of the
  sample. This figure is identical to Figure 18 of BC03.
}
\end{figure}

%% file: fig7.tex
\begin{figure}
 \setbox20=\hbox
 {\psfig{file=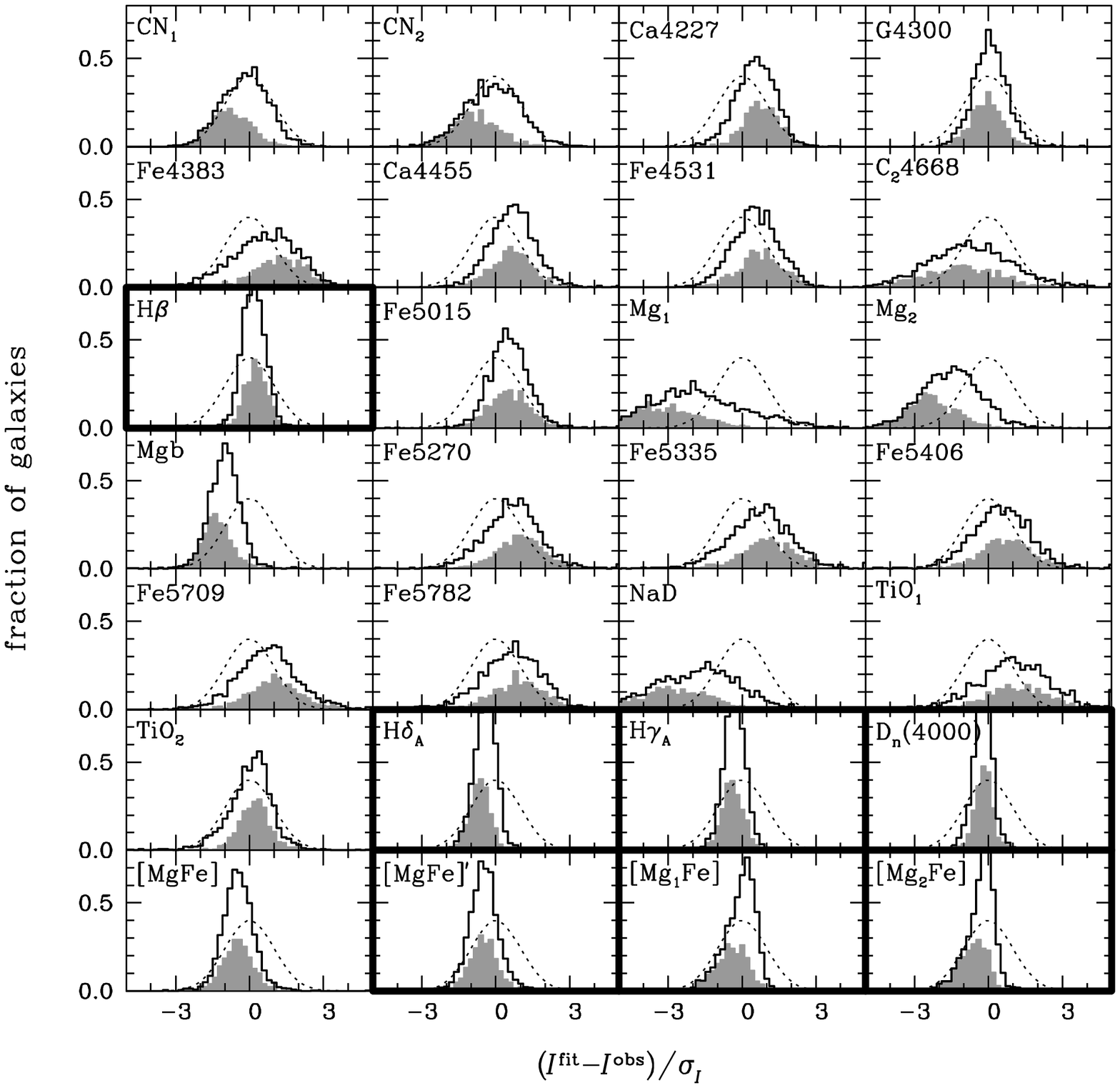,width=14cm,height=20cm,angle=0}}
 \centerline{\box20}
 \caption{
  Simultaneous fit of the strengths of several indices in the spectra
  of 2010 galaxies with S/N$_{\rm med} \ge 30$ in the `main galaxy sample'
  of the SDSS EDR using BC03 models built with the IndoUS library.
  The highlighted frames indicate the seven indices used
  to constrain the fits. The stellar velocity dispersion of the models are
  required to be within 15 km s$^{-1}$ of the observed ones. Each panel shows
  the distribution of the fitted index strength $I^{\rm fit}$ minus the
  observed one $I^{\rm obs}$, divided by the associated error $\sigma_I$.
  For reference, a dotted line in each panel indicates a Gaussian distribution
  with unit standard deviation. The shaded histograms show the contributions
  to the total distributions by galaxies with $\sigma_V >180$ km s$^{-1}$,
  corresponding roughly to the median stellar velocity dispersion of the
  sample.
}
\end{figure}

%% file: fig8.tex
\begin{figure}
 \setbox20=\hbox
 {\psfig{file=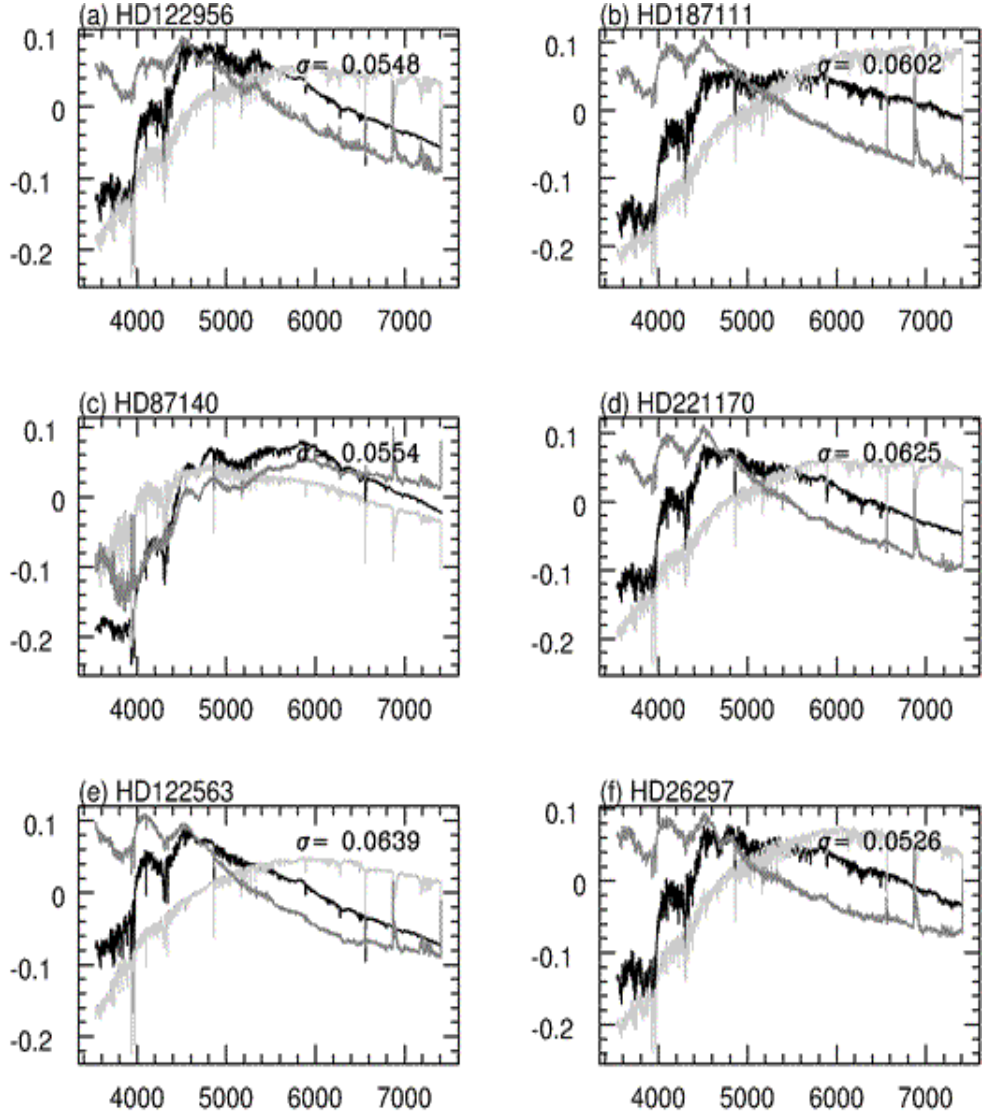,width=14cm,height=20cm,angle=0}}
 \centerline{\box20}
 \vskip -4.00cm
 \caption{
  Comparison of the sed's of 6 stars present in the IndoUS and the MILES
  stellar libraries in the wavelength range $\lambda\lambda$ 3500 - 7400 \AA\
  in common in these libraries. In each panel the black line is the IndoUS
  sed, the light gray line is the MILES sed, and the dark gray line represents
  the difference (IndoUS - MILES). The sed's have been scaled arbitrarily
  in the flux scale. The star identification is indicated in the upper
  left corner of each panel.
}
\end{figure}

%% file: fig9.tex
\begin{figure}
 \setbox20=\hbox
 {\psfig{file=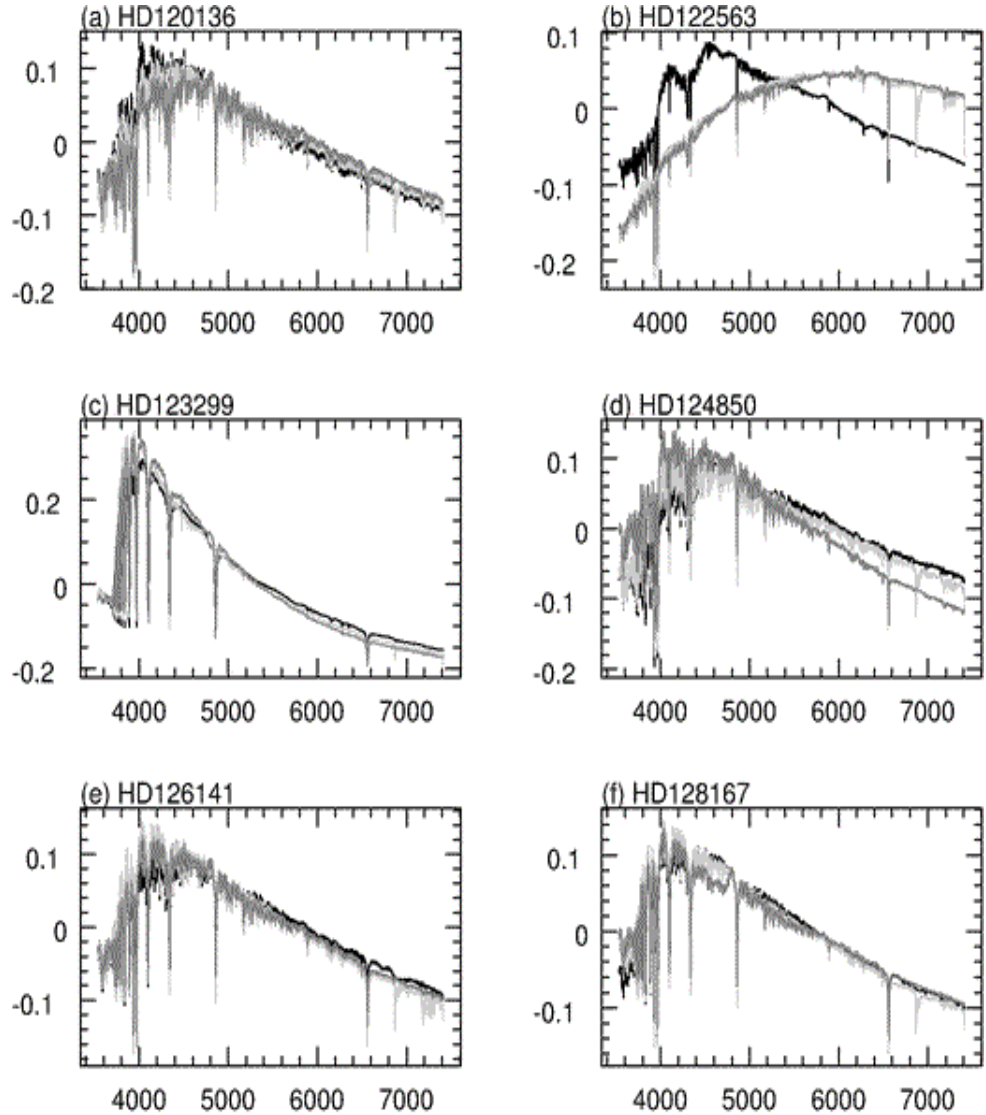,width=14cm,height=20cm,angle=0}}
 \centerline{\box20}
 \vskip -4.00cm
 \caption{
  Comparison of the sed's of 6 stars present in the IndoUS, the STELIB, and
  the MILES stellar libraries in the wavelength range $\lambda\lambda$
  3500 - 7400 \AA\ in common in these libraries. In each panel the black line
  is the IndoUS sed, the light gray line is the MILES sed, and the dark gray
  the STELIB sed. The sed's have been scaled arbitrarily in the flux scale.
  The star identification is indicated in the upper left corner of each panel.
}
\end{figure}

%% file: fig10.tex
\begin{figure}
 \setbox20=\hbox
 {\psfig{file=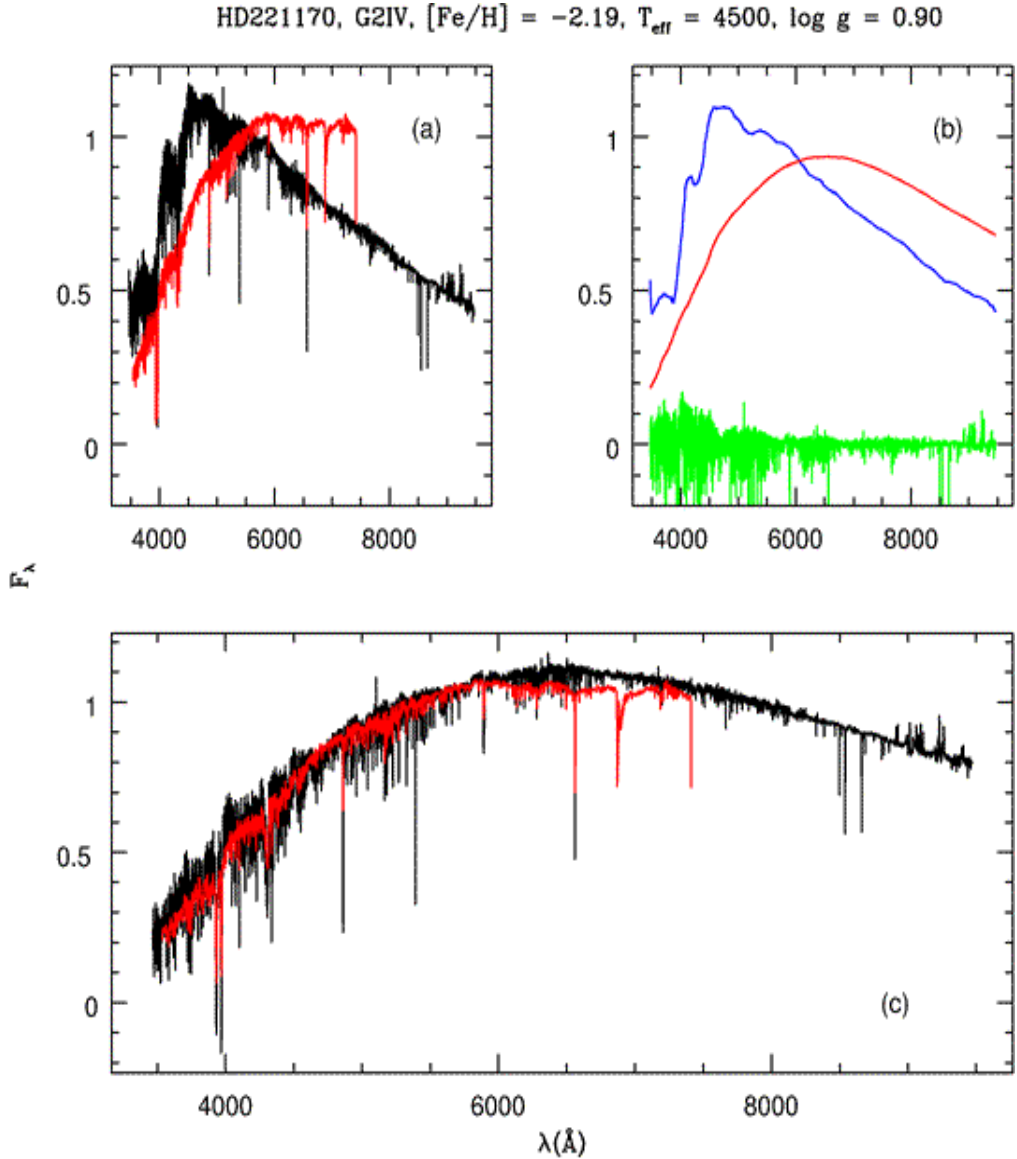,width=14cm,height=20cm,angle=0}}
 \centerline{\box20}
 \vskip -4.00cm
 \caption{
  Spectral energy distributions of the star HD221170.
  {\it (a).} The MILES sed extends up to 7400 \AA\ and the IndoUS sed up to
  9500 \AA.
  {\it (b).} Smoothed version of the IndoUS sed for this star and of the
  BaSeL 3.1 model for $T_{eff}$ = 4500K, log g = 1, and [Fe/H] = -2,
  corresponding to the tabulated values for HD221170. The residual line in the
  bottom part of the panel is the difference between the original and
  the smoothed IndoUS sed's.
  {\it (c).} Result of adding the smoothed BaSeL sed to the difference of the
  original and smoothed IndoUS sed's shown in frame {\it (b)}.
}
\end{figure}

%% file: fig11.tex
\begin{figure}
 \setbox20=\hbox
 {\psfig{file=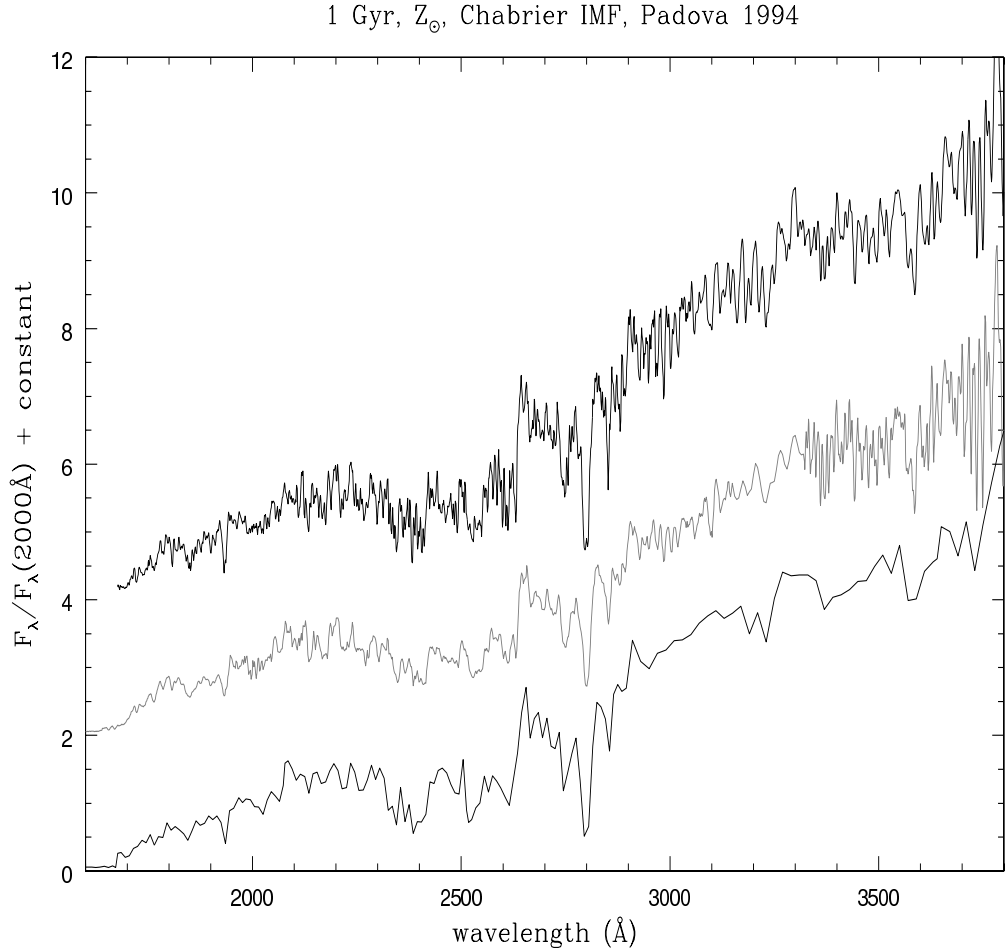,width=14cm,height=13cm,angle=270}}
 \centerline{\box20}
 \caption{
  BC03 standard SSP model spectra for solar metallicity at 1 Gyr in
  the wavelength range from 1600 \AA\ to 3800 \AA. The spectrum on top
  (thick black line) is built using the HNGSL, the one in the middle
  (gray line) uses the Pickles library and STELIB, and the bottom one
  (thin black line) the BaSel 3.1 library. The spectra have been
  normalized at 2000 \AA\ and shifted arbitrarily in the vertical
  direction for clarity.
  }
\end{figure}

%% file: fig12.tex
\begin{figure}
 \setbox20=\hbox
 {\psfig{file=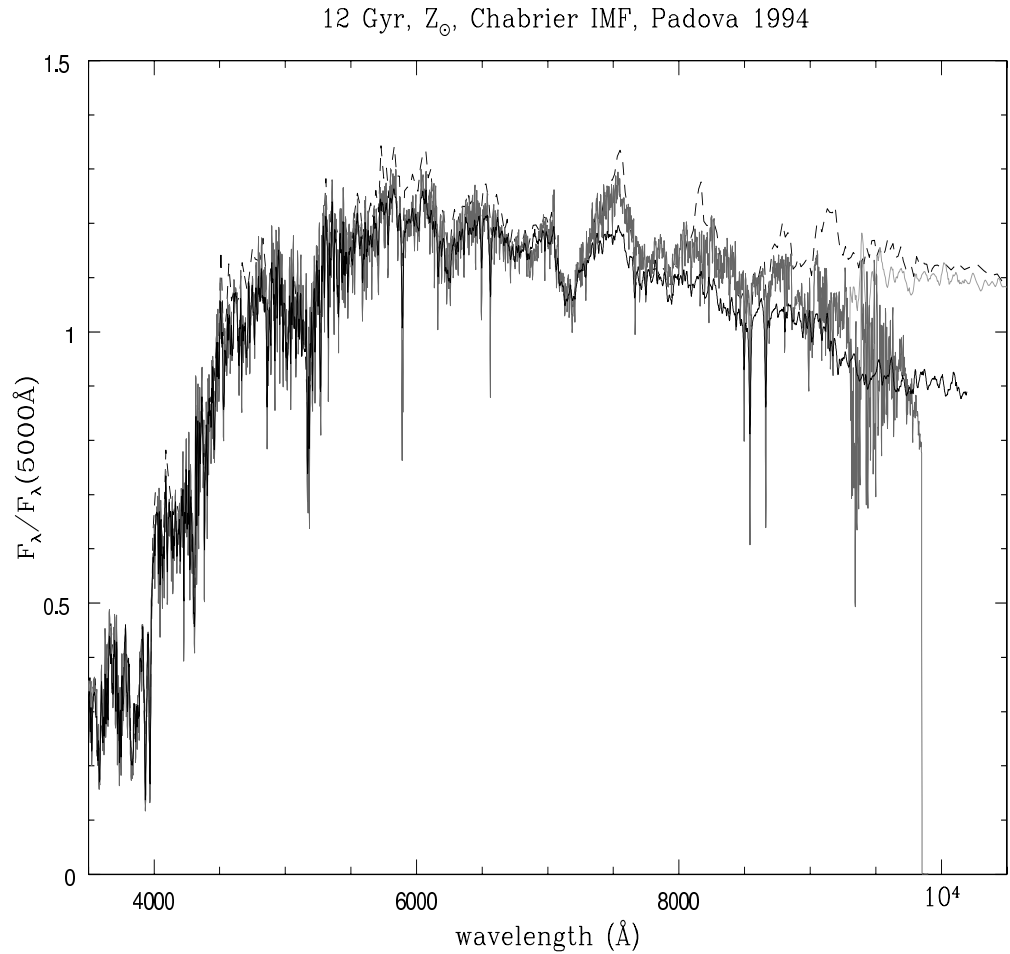,width=14cm,height=13cm,angle=270}}
 \centerline{\box20}
 \caption{
  BC03 standard SSP model spectra for solar metallicity at 12 Gyr in
  the wavelength range from 3500 \AA\ to 10,500 \AA.
  The black solid line (reaching up to 10,200 \AA) represents the
  spectrum built using the HNGSL,
  the dark gray solid line (noisy above 9000 \AA) uses STELIB,
  the light gray solid line uses STELIB up to 9000 \AA\ and the
  Pickles library at longer wavelengths,
  the black dashed line uses the BaSel 3.1 library.
  The spectra have been normalized at 5000 \AA.}
\end{figure}

%% file: fig13.tex
\begin{figure}
 \setbox20=\hbox
 {\psfig{file=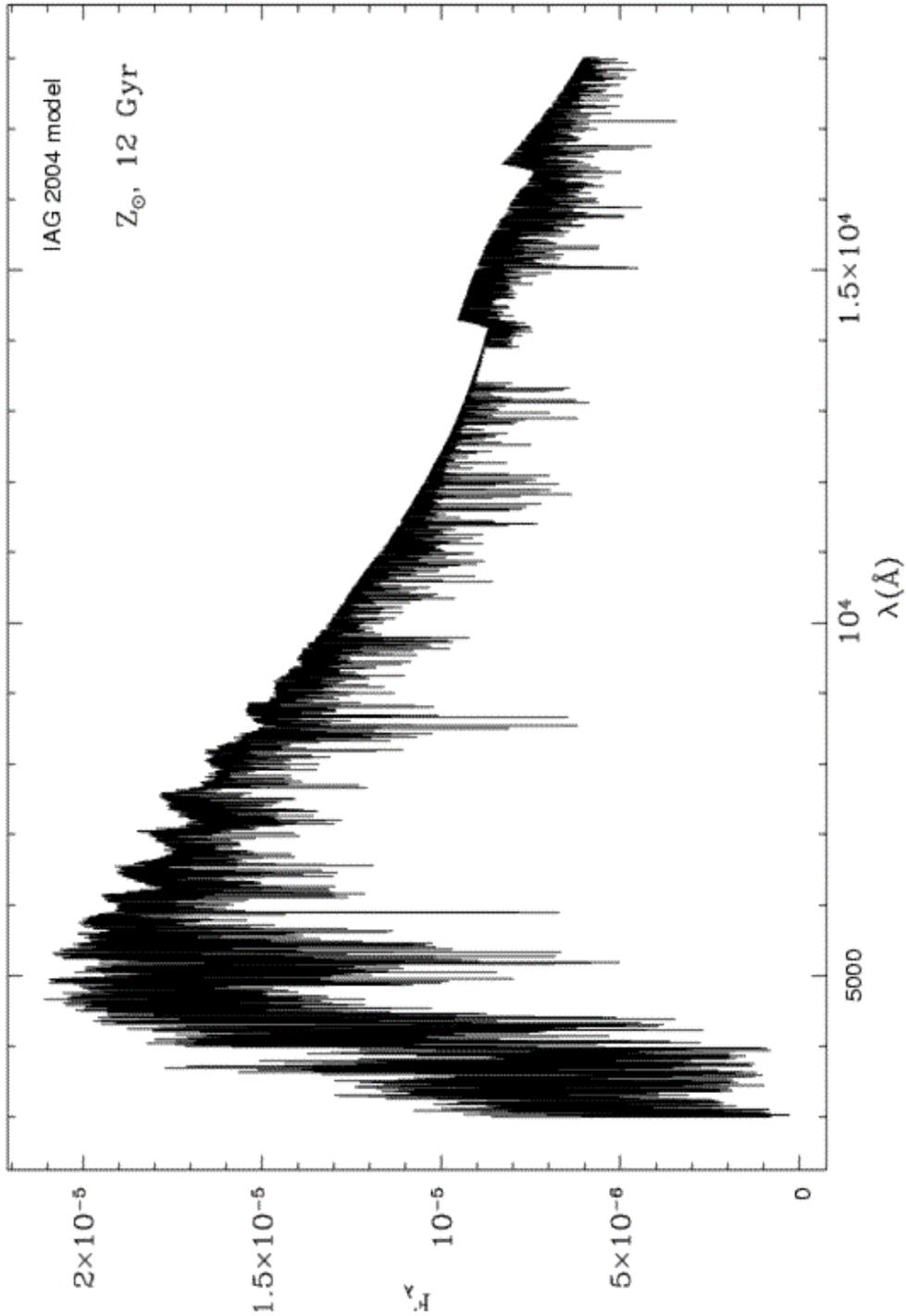,width=14cm,height=20cm,angle=0}}
 \centerline{\box20}
 \caption{
  BC03 standard SSP model sed at 12 Gyr computed with the high resolution
  theoretical model atmospheres from the IAG collaboration for [Fe/H] = 0,
  $[\alpha/Fe] = 0$ \citep{COE05} in the wavelength range from 0.3 to
  18 $\mu$m.
}
\end{figure}

%% file: fig14.tex
\begin{figure}
 \setbox20=\hbox
 {\psfig{file=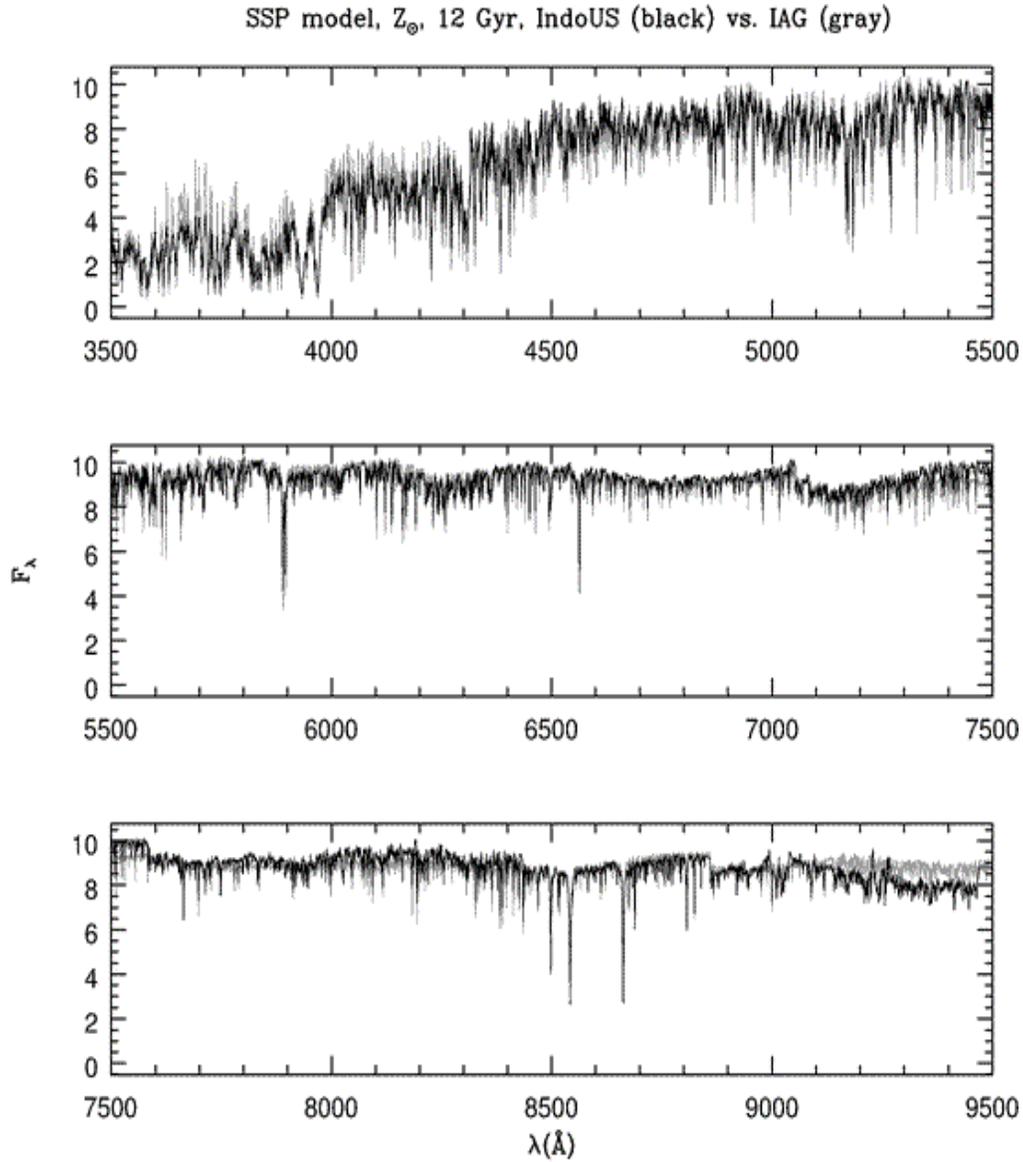,width=14cm,height=20cm,angle=0}}
 \centerline{\box20}
 \vskip -4.00cm
 \caption{
  BC03 standard SSP model sed at 12 Gyr computed with the high resolution
  theoretical model atmospheres from the IAG collaboration for [Fe/H] = 0,
  $[\alpha/Fe] = 0$ \citep{COE05} in the wavelength range from 3500 to 9500
  \AA\ (gray line) downgraded to the spectral resolution of the IndoUS library.
  For comparison, the corresponding model computed with the IndoUS stellar library
  is shown as a black line.
}
\end{figure}

%% file: fig15.tex
\begin{figure}
 \setbox20=\hbox
 {\psfig{file=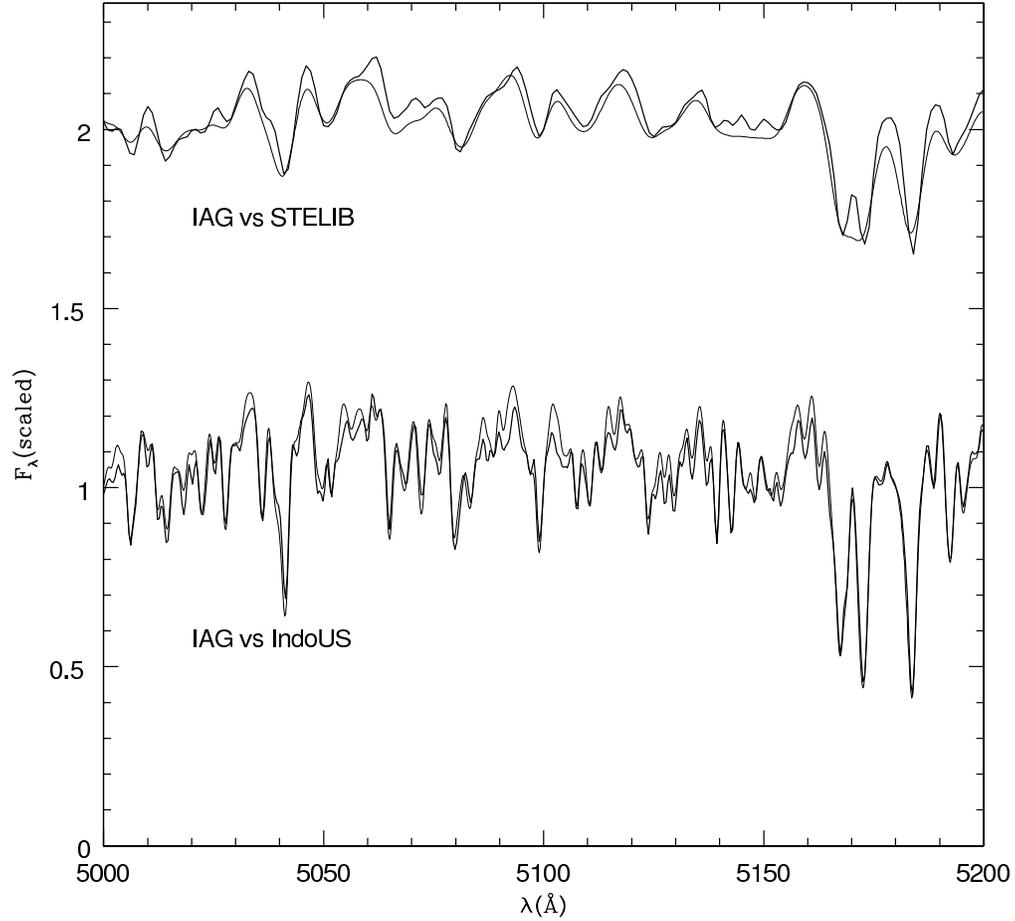,width=14cm,height=16cm,angle=0}}
 \centerline{\box20}
 \caption{
  BC03 standard SSP model sed at 12 Gyr computed with the high resolution theoretical
  model atmospheres from the IAG collaboration for [Fe/H] = 0, $[\alpha/Fe] = 0.0$
  \citep{COE05}, in the wavelength range from 5000 to 5200 \AA\ containing the
  Mg$_1$, Mg$_2$, and Mgb features. The thin lines show the IAG model downgraded
  to the spectral resolution of the STELIB and IndoUS libraries, shown as heavy
  lines. The spectra have been scaled and shifted arbitrarily in the vertical
  direction.
}
\end{figure}

%% file: fig16.tex
\begin{figure}
 \setbox20=\hbox
 {\psfig{file=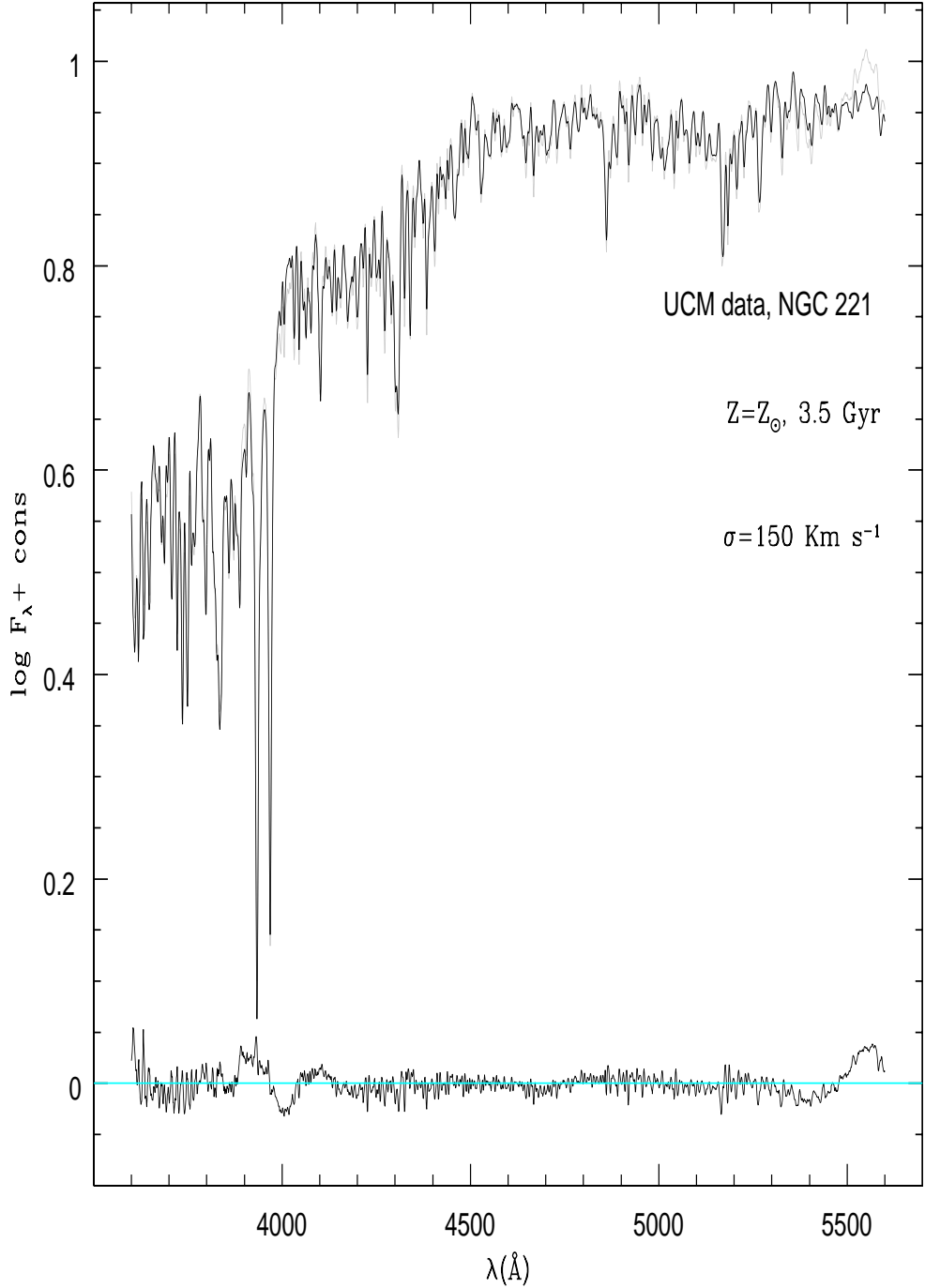,width=14cm,height=20cm,angle=0}}
 \centerline{\box20}
 \caption{
  Best SSP model fit to the spectrum of the elliptical galaxy NGC 221 in the
  wavelength range from 3600 to 5600 \AA.
  The BC03 standard SSP model computed with the IndoUS stellar library for
  solar metallicity at an age of 3.5 Gyr is shown in black. The stellar velocity
  dispersion $\sigma = 150$ km s$^{-1}$ applied to the model spectrum is
  also derived by the fitting algorithm. The observed spectrum is shown in
  gray and was kindly provided by P. S\'anchez-Bl\'azquez.
  The residuals (observed - model) are shown at the bottom of the plot in the
  same vertical scale.
}
\end{figure}

%% file: fig17.tex
\begin{figure}
 \setbox20=\hbox
 {\psfig{file=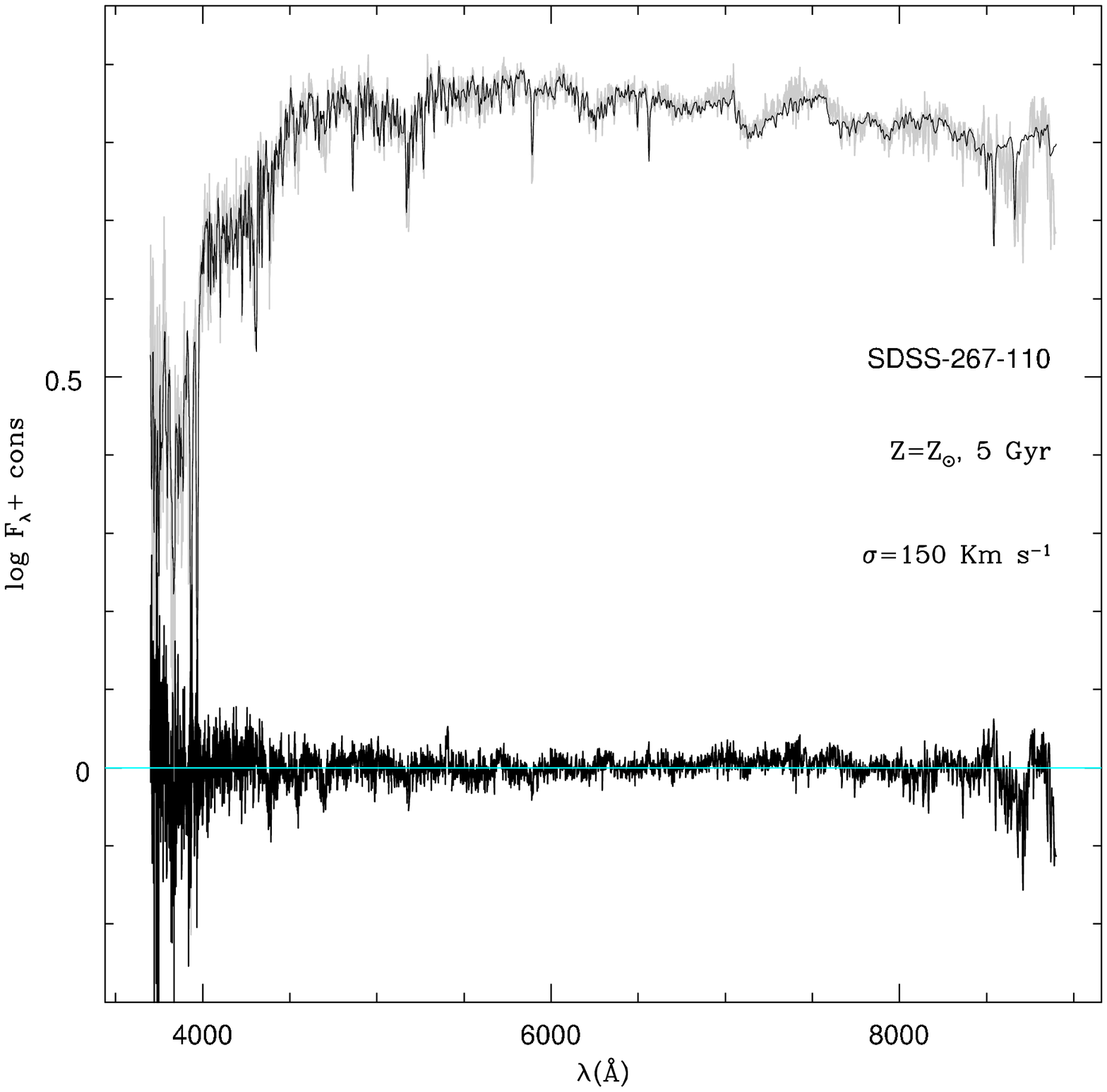,width=14cm,height=20cm,angle=0}}
 \centerline{\box20}
 \vskip -4.00cm
 \caption{
  Best SSP model fit to the spectrum of the early-type galaxy SDSS-267-110
  in the wavelength range from 3750 to 8800 \AA.
  The BC03 standard SSP model computed with the IndoUS stellar library for
  solar metallicity at an age of 5 Gyr is shown in black. The stellar velocity
  dispersion $\sigma = 150$ km s$^{-1}$ applied to the model spectrum is also
  derived by the fitting algorithm. The SDSS-EDR observed spectrum is shown
  in gray.
  The residuals (observed - model) are shown at the bottom of the plot in the
  same vertical scale.
}
\end{figure}

%% file: fig18.tex
\begin{figure}
 \setbox20=\hbox
 {\psfig{file=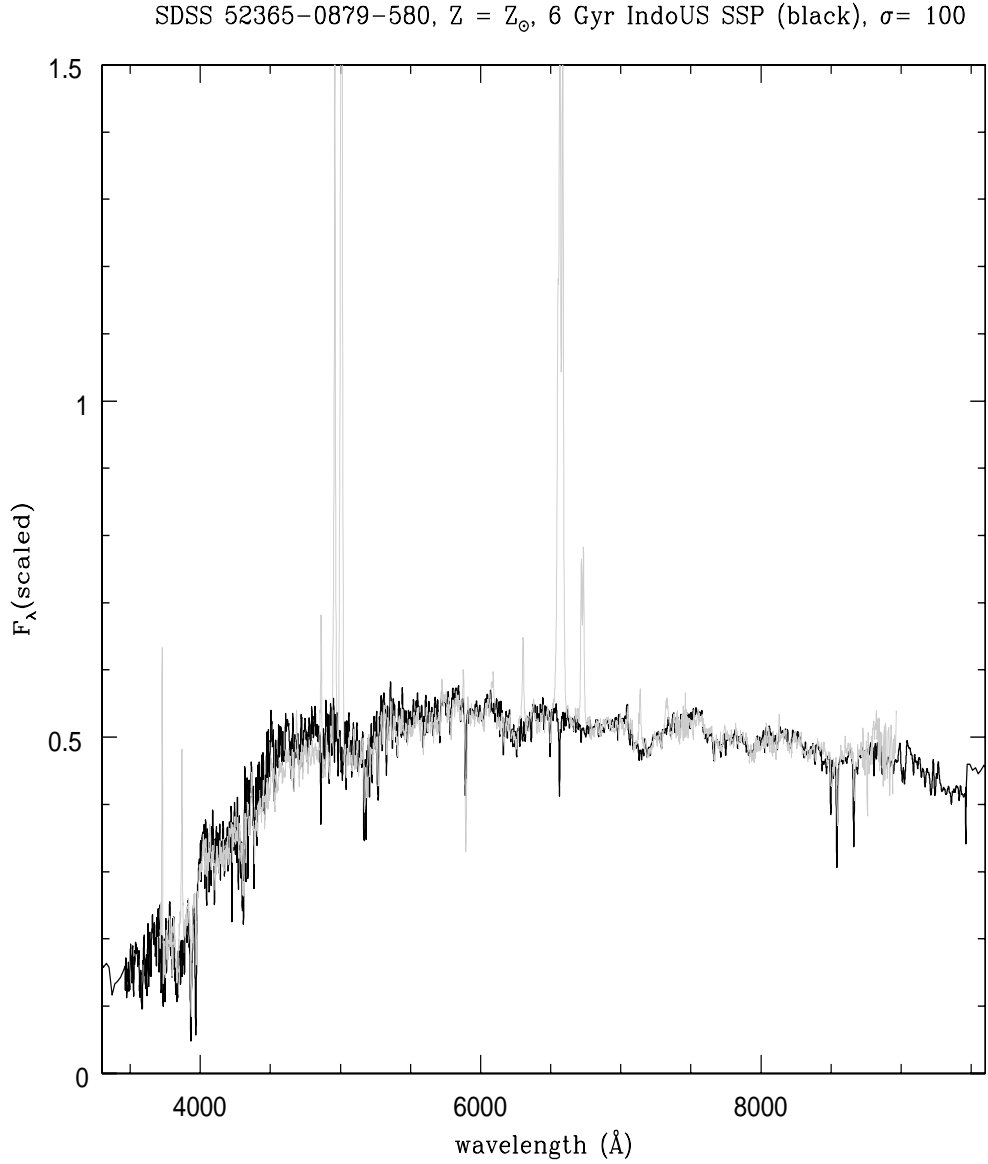,width=14cm,height=16cm,angle=0}}
 \centerline{\box20}
 \caption{
  Best SSP model fit to the continuum spectrum of the star forming galaxy
  SDSS-52365-0879-580 in the wavelength range from 3500 to 9500 \AA.
  The BC03 standard SSP model computed with the IndoUS stellar library for
  solar metallicity at an age of 6 Gyr is shown in black. The stellar velocity
  dispersion $\sigma = 100$ km s$^{-1}$ applied to the model spectrum is also
  derived by the fitting algorithm. The SDSS-EDR observed spectrum is shown
  in gray. The wavelength ranges containing emission lines in the observed
  spectrum are not included in the fit.
}
\end{figure}

%% file: fig19.tex
\begin{figure}
 \setbox20=\hbox
 {\psfig{file=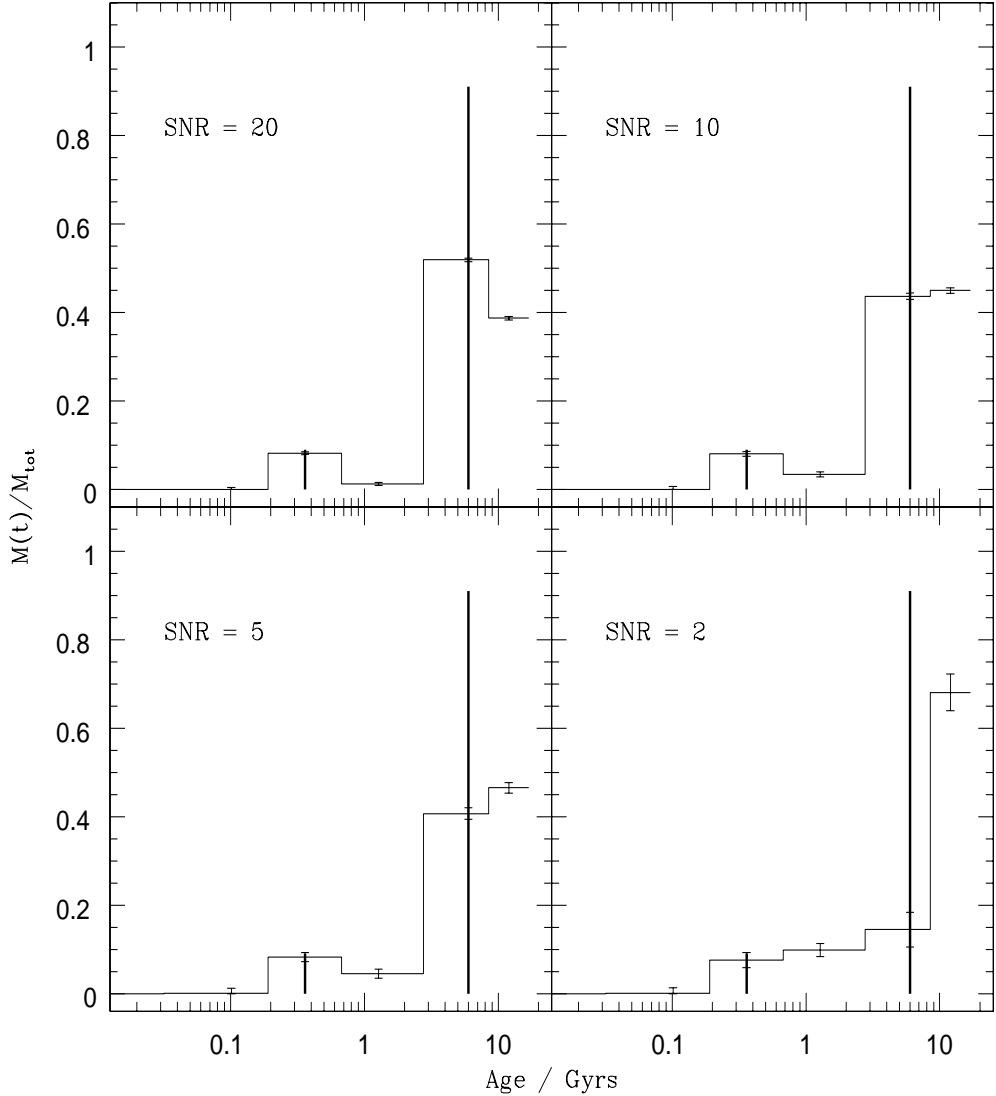,width=14cm,height=16cm,angle=0}}
 \centerline{\box20}
 \caption{
  Recovery of a known star formation history as a function of the
  signal-to-noise ratio of the problem spectrum. The problem spectrum
  corresponds to an old stellar population of age 6 Gyr in which a second
  burst of star formation of $1 \over 9$ the intensity of the first burst
  occurs 360 Myr ago. The two vertical lines at the corresponding ages
  represent schematically this star formation history. The histograms
  with error bars represent the star formation history recovered by the
  GASPEX non-parametric CSP fitting algorithm described by \citet{MAT05} when noise
  is added to the problem spectrum. The value of the signal-to-noise ratio
  (SNR) used is indicated in each frame.
}
\end{figure}

%% file: fig20.tex
\begin{figure}
 \setbox20=\hbox
 {\psfig{file=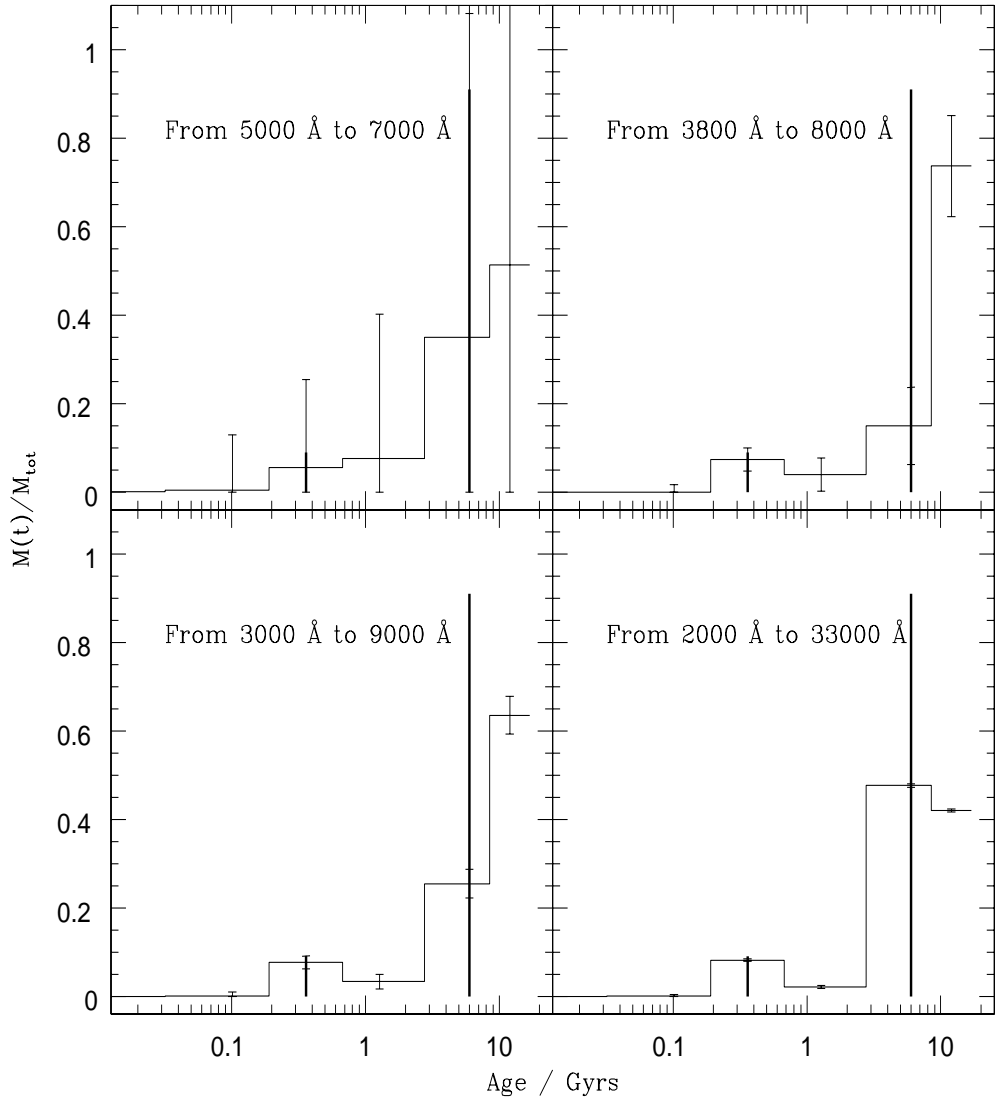,width=14cm,height=16cm,angle=0}}
 \centerline{\box20}
 \caption{
  Recovery of a known star formation history as a function of
  wavelength coverage. The problem spectrum is the same one used in
  Figure 19 for a signal-to-noise ratio of 20.
  The histograms with error bars represent the star formation history
  recovered by the GASPEX non-parametric CSP fitting algorithm described by
  \citet{MAT05} as a function of the wavelength range used in the fit, as
  indicated in each frame.
}
\end{figure}

%% file: fig21.tex
\begin{figure}
 \setbox20=\hbox
 {\psfig{file=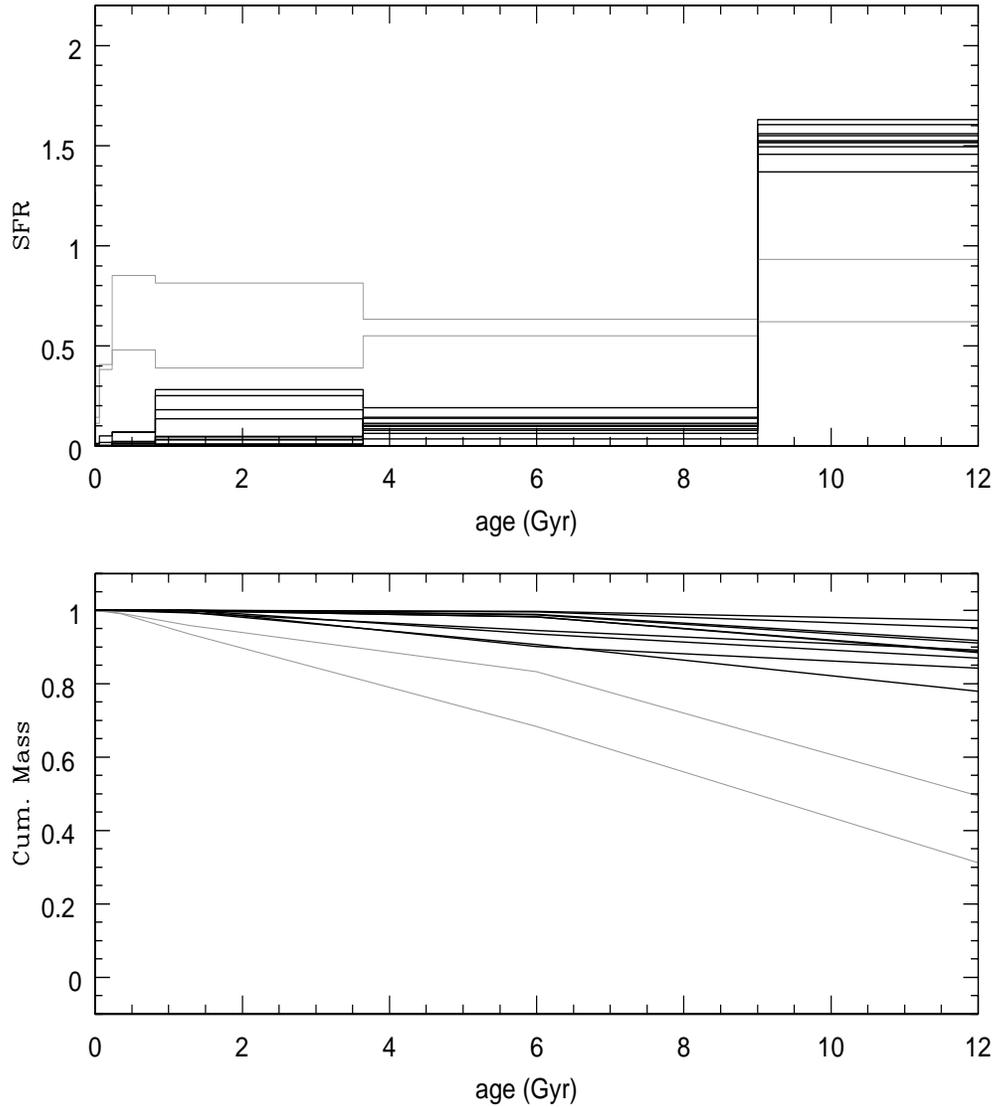,width=14cm,height=16cm,angle=0}}
 \centerline{\box20}
 \caption{
  Star formation history and cumulative mass vs. age recovered by the
  GASPEX non-parametric CSP fitting algorithm described by Mateu, Magris, \&
  Bruzual (2005) for a
  selection of galaxies from the Near Field Galaxy Survey (Jansen et al. 2000).
  The gray lines represent galaxies with significant recent star formation,
  as indicated by the emission lines in their spectra.
}
\end{figure}

%% file: fig22.tex
\begin{figure}
 \setbox20=\hbox
 {\psfig{file=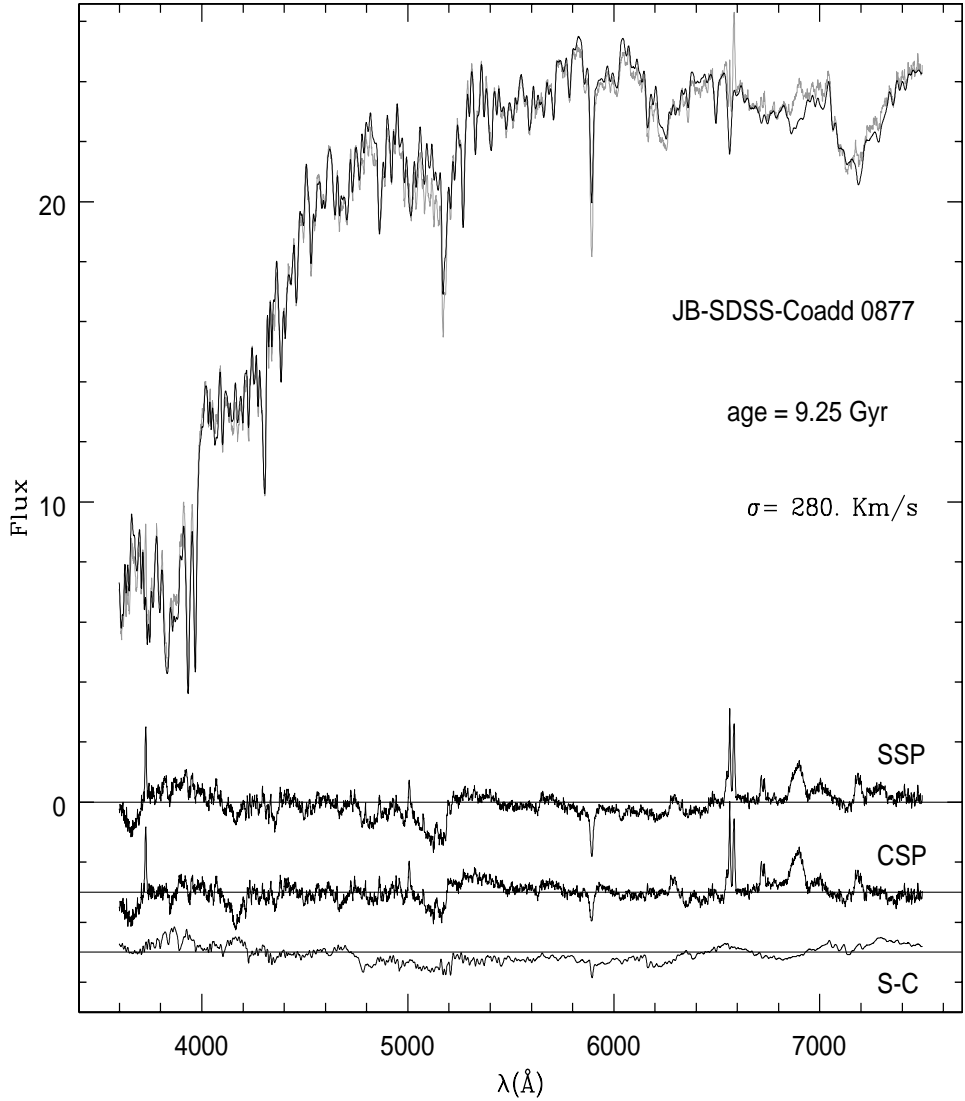,width=14cm,height=16cm,angle=0}}
 \centerline{\box20}
 \caption{
  Best model fit to the spectrum of an early type galaxy resulting from
  co-adding the spectra of several galaxies from the SDSS until reaching
  a signal-to-noise ratio of 250. This spectrum, kindly
  provided by Jarle Brinchmann, is shown as a light line in the wavelength
  range from 3600 to 7500 \AA.
  The BC03 standard SSP model computed with the STELIB stellar library for
  solar metallicity at an age of 9.25 Gyr is shown as a heavy line.
  The stellar velocity dispersion $\sigma = 280$ km s$^{-1}$ applied to the
  model spectrum is also derived by the fitting algorithm.
  At the bottom of the figure the line marked SSP represents the residuals
  (observed -model) for the SSP model shown in the figure.
  The line marked CSP represents the residuals obtained when the fit is
  performed by the GASPEX non-parametric CSP fitting algorithm described by
  Mateu, Magris, \& Bruzual (2005).
  The line marked S-C shows the difference between the SSP and CSP solutions.
  The last two lines have been shifted down for clarity.
}
\end{figure}

%% file: fig23.tex
\begin{figure}
 \setbox20=\hbox
 {\psfig{file=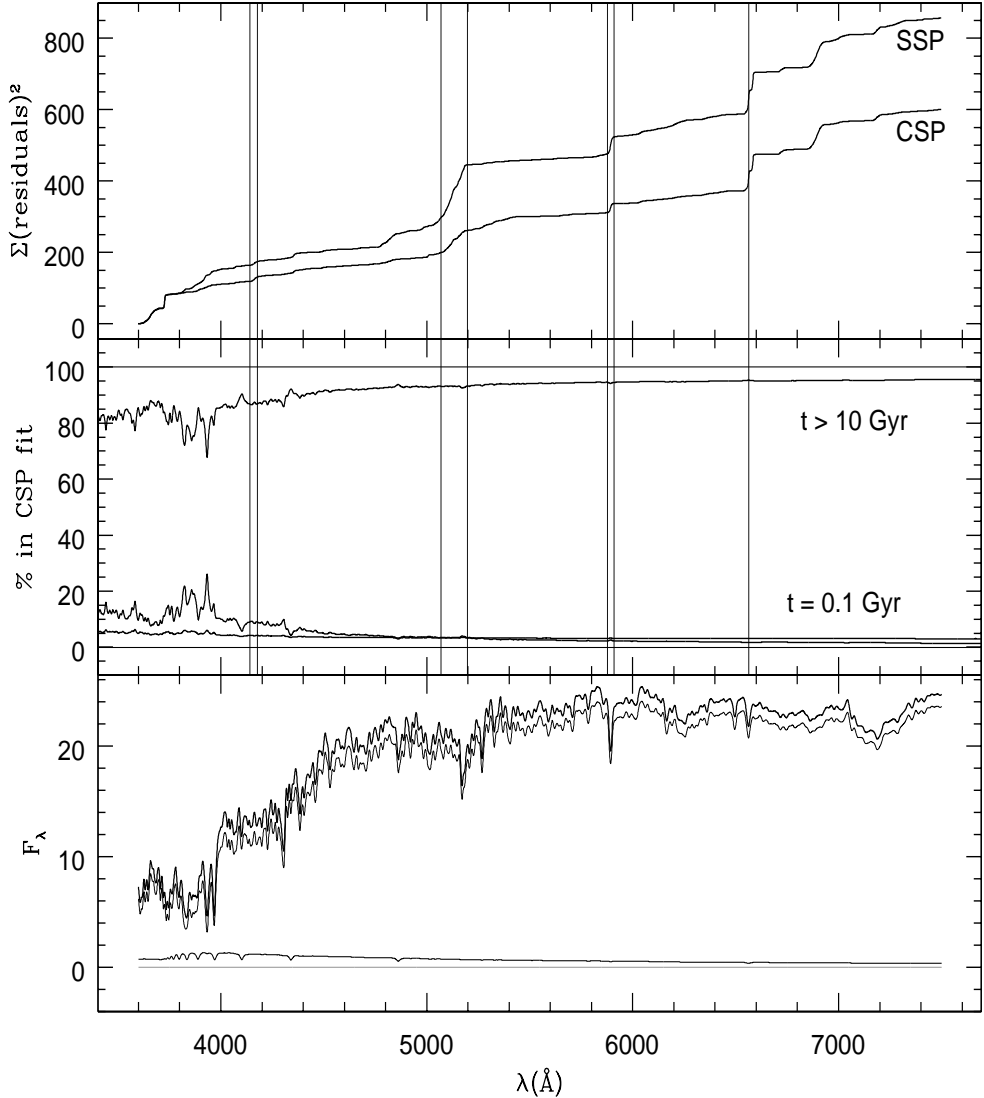,width=14cm,height=16cm,angle=0}}
 \centerline{\box20}
 \caption{
  {\it Top frame.} Cumulative squared residuals for the SSP and CSP
  fits shown in Figure 22.
  The vertical lines show the position of the central bands defining
  the CN$_1$-CN$_2$, Mg$_1$-Mg$_2$-Mgb, and NaD Lick indices, and the
  H$\alpha$ line.
  {\it Middle frame}. Percentage contribution of the old and a very
  young population (100 Myr) to the total spectrum of this galaxy in
  the CSP solution as a function of wavelength. The contribution of
  populations of other ages included in the GASPEX solution is, added
  all together, below the 5\% level at $\lambda < 4500$ \AA, and is 
  even less at longer wavelengths.
  {\it Bottom frame}. Spectra corresponding to the old, young, and
  the rest of the stellar populations in the GASPEX solution.
  The latter contribute essentially zero flux to the total sed,
  represented by the heavy line.
}
\end{figure}

%% file: fig24.tex
\begin{figure}
 \setbox20=\hbox
 {\psfig{file=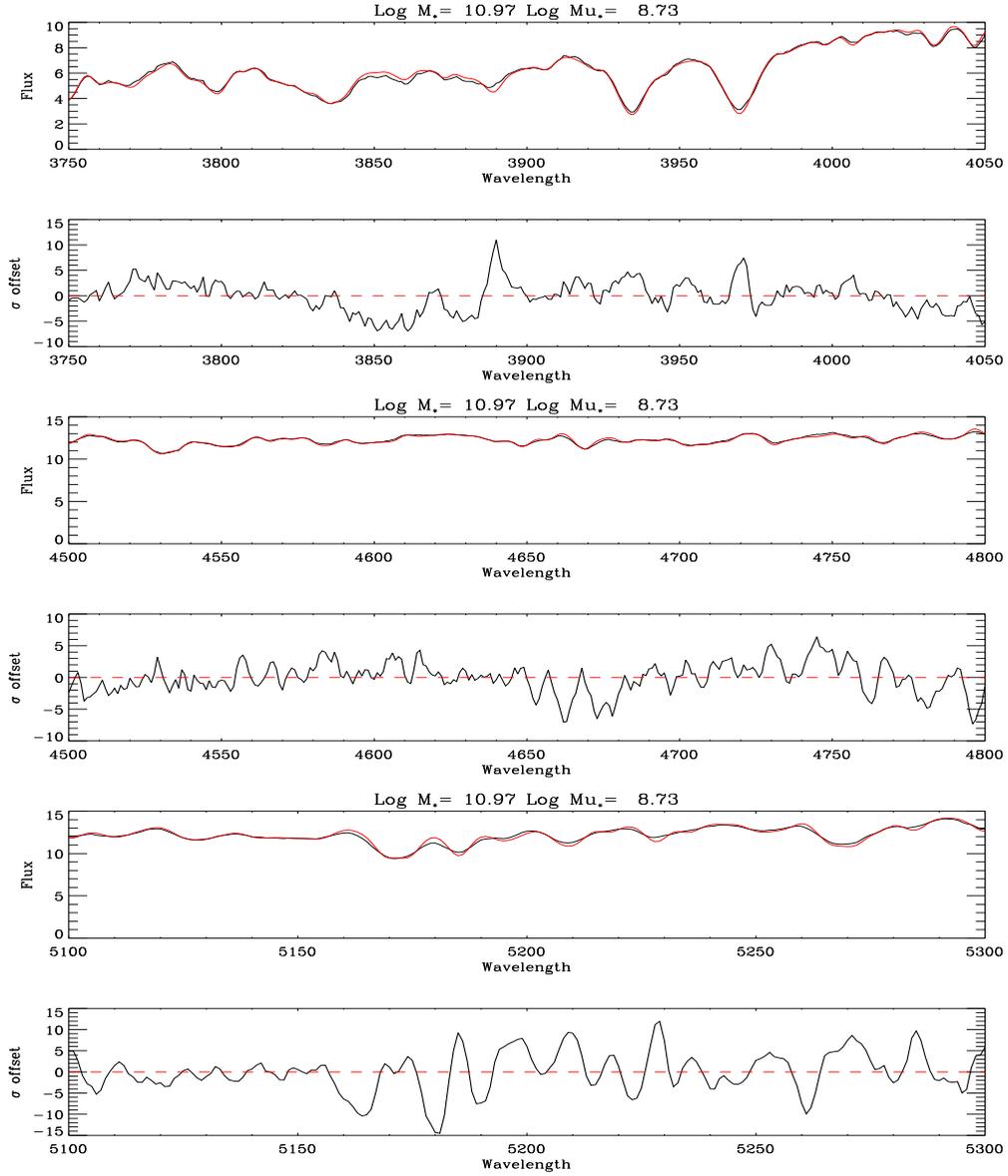,width=14cm,height=16cm,angle=0}}
 \centerline{\box20}
 \caption{
  Detailed comparison of the fit to one of J. Brinchmann co-added
  spectrum in specific wavelength regions. The residuals in the second,
  fourth and sixth panels are expressed in units of the standard deviation
  $\sigma$ of the co-added fluxes.
}
\end{figure}